\documentclass[a4paper]{article}
\usepackage[T1]{fontenc}
\usepackage{graphicx}
\usepackage{url}
\usepackage{amsfonts,amsmath,amssymb,lscape,float}
\usepackage{bm,dsfont}
\usepackage{xcolor}
\usepackage{bbm}
\usepackage{braket}
\usepackage{longtable}
\usepackage[dvips]{epsfig}
\usepackage[colorlinks=true,linkcolor=blue, allcolors=blue]{hyperref}
\usepackage{listings}
\usepackage{etoolbox}
\usepackage{authblk}
\usepackage{subcaption}
\usepackage{mathtools}

\makeatletter
\newcommand{\phantomlabel}[2]{
    \protected@write\@auxout{}{
        \string\newlabel{#2}{
            {\@currentlabel#1}{\thepage}
            {\@currentlabel#1}{#2}{}
        }
    }
    \hypertarget{#2}{}
}

\makeatletter
\def\@fnsymbol#1{\ensuremath{\ifcase#1\or \dagger\or \ddagger\or
   \mathsection\or \mathparagraph\or \|\or **\or \dagger\dagger
   \or \ddagger\ddagger \else\@ctrerr\fi}}
    \makeatother

\makeatother

\usepackage[sorting=none, giveninits=true, style=nature, url=false, doi=false]{biblatex}
\bibliography{refs}

\begin{document}
\title{Mapping quantum circuits to shallow-depth measurement patterns based on graph states} 

\def\thefootnote{*}\footnotetext{Both authors contributed equally to this work.}

\author[a,c]{Thierry N. Kaldenbach *\thanks{\textcolor{blue}{thierry.kaldenbach@dlr.de}}}
\author[a,b]{Matthias Heller *\thanks{\textcolor{blue}{matthias.heller@igd.fraunhofer.de}}}
\affil[a]{Fraunhofer Institute for Computer Graphics Research IGD, Darmstadt, Germany}
\affil[b]{Technical University of Darmstadt, Interactive Graphics Systems Group, Germany
}
\affil[c]{German Aerospace Center (DLR), Institute of Materials Research, Cologne, Germany}

\date{}
\maketitle

\begin{abstract}
    The paradigm of measurement-based quantum computing (MBQC) starts from a highly entangled resource state on which unitary operations are executed through adaptive measurements and corrections ensuring determinism. This is set in contrast to the more common quantum circuit model, in which unitary operations are directly implemented through quantum gates prior to final measurements. In this work, we incorporate concepts from MBQC into the circuit model to create a hybrid simulation technique, permitting us to split any quantum circuit into a classically efficiently simulatable Clifford-part and a second part consisting of a stabilizer state and local (adaptive) measurement instructions -- a so-called standard form -- which is executed on a quantum computer. We further process the stabilizer state with the graph state formalism, thus enabling a significant decrease in circuit depth for certain applications. We show that groups of fully commuting operators can be implemented using fully-parallel, i.e., non-adaptive, measurements within our protocol. In addition, we discuss how such circuits can be implemented in constant quantum depths by employing quantum teleportation. Finally, we demonstrate the utility of our technique on two examples of high practical relevance -- the Quantum Approximate Optimization Algorithm (QAOA) and the Variational Quantum Eigensolver (VQE).
\end{abstract}

\clearpage

\section{Introduction}

Measurement-based quantum computing (MBQC) offers an interesting alternative model of quantum computing compared to the standard circuit model.
While unitary operations in the circuit model are realized by sequential application of quantum gates, MBQC operates on a highly entangled state, called resource state, on which unitary operations can be implemented via adaptive measurements \cite{raussendorf2001one, jozsa2006introduction,PhysRevLett.86.910}.
Some of these measurements can be performed in parallel, which leads to a compelling feature of MBQC: the parallel application of unitaries, which in the gate model would be applied sequentially. 

Due to its universality, it is possible to map any quantum circuit to the MBQC model and vice versa.
Forward- and backward translation between the circuit model and MBQC can lead to beneficial tradeoffs in terms of depth and space complexity \cite{broadbent2009parallelizing,PhysRevLett.126.220501}. 
Different techniques for translation between these models \cite{miyazaki2015analysis, backens2021there} and optimization of adaptive measurement patterns \cite{eslamy2018optimization} have been studied based on graph-theoretical tools such as causal flow \cite{PhysRevA.74.052310} and its generalization~\cite{browne2007generalized}.

In this article, we introduce a straight-forward algorithm that allows the mapping of a given quantum circuit to a graph state, which, together with local Clifford operations and measurement instructions, allows one to perform quantum computations within the MBQC model. 
A closely related circuit transpilation approach has recently been explored in Ref.~\cite{vijayan2022compilation}, where circuits are first transformed into an inverse Initialization-CNOT-Measurement (ICM) form \cite{herr2017lattice, paler2014mapping, paler2017fault} via $T$-state injection \cite{nielsen2002quantum}, followed by gate teleportation of $T$-gates to separate the Clifford structure, which is then mapped to a graph state.  
For our algorithm, the point of departure is not an inverse ICM form, but instead a quantum circuit generated by a sequence of multi-qubit Pauli exponentials.
Since any quantum circuit can be expressed in terms of such exponentials, our framework is general.
The same circuit structure has been considered in Ref.~\cite{chan2023hybrid}, in which the Pauli exponentials were called gadgets.

Each Pauli exponential has precisely one non-Clifford gate, which is the $R_z$-rotation gate. 
We show how this gate can be implemented through gate teleportation, which is similar to the treatment of $T$-gates in Ref.~\cite{vijayan2022compilation}.
This allows us to derive measurement patterns based on graph states, which implement the initial quantum circuit. We employ the simulated annealing algorithm \cite{van1987simulated} to minimize the preparation cost of the graph states on real hardware by exploiting the local Clifford equivalence between different graph states. 

Being able to map circuits to graph states and vice versa, we show how our algorithm can be used to parallelize the application of $n$ commuting Pauli exponentials at the expense of introducing $n$ ancilla qubits.
As we will see, this follows from the standard form~\cite{broadbent2009parallelizing, danos2007measurement} of the measurement patterns we derive: any pattern can be decomposed into a Clifford-part, an adaptive measurement part and finally a corrective Pauli layer. 
The Clifford part can be implemented in constant quantum depth using quantum teleportation for parallelization \cite{jozsa2006introduction}, whereas the depth of the adaptive measurement part scales linearly with the number of fully-commuting groups of the generator of the initial gate-based circuit.

We use our algorithm in two scenarios, which are often discussed as near-term applications for noisy intermediate-scale quantum (NISQ)-devices: the variational Quantum Eigensolver (VQE) \cite{peruzzo2014variational, TILLY20221} in the context of molecular simulations and the Quantum Approximate Optimization Algorithm (QAOA) \cite{farhi2014quantum, zhou2020quantum} in the context of binary optimization problems. 
In particular, for the ground-state energy calculation of H\textsubscript{2}O, we demonstrate that our algorithm achieves an efficient mapping of the Qubit-ADAPT-VQE~\cite{PRXQuantum.2.020310} to highly shallow circuits, thus advancing the practical utility of quantum computation in the NISQ-era \cite{preskill2018quantum} where circuit depth is strictly limited by coherence time. 

The remainder of this article is structured as follows: Sec.~\ref{sec:MBQC} provides an overview on the graph state formalism (Sec.~\ref{sec:GraphStates}) and MBQC (Sec.~\ref{sec:MBQC2}). 
Based on these fundamentals, we proceed to introduce our algorithm step by step in Sec.~\ref{sec:MappingGraphStates}. 
We start by reviewing common circuit structures for Pauli exponentials \cite{nielsen2002quantum, gui2020term} and combine them with the One-Way Quantum Computer (1WQC) protocol \cite{raussendorf2001one, raussendorf2001computational, PhysRevA.68.022312, raussendorf2009measurement} in Sec.~\ref{sec:Replace}, revealing a general circuit structure consisting of Clifford-, measurement- and Pauli parts. 
In Sec.~\ref{sec:Parallelism}, we elaborate how commuting Pauli strings can be implemented through parallel measurements. Sec.~\ref{sec:ParallelClifford} details how the Clifford structure may be implemented with constant circuit depth. Next, in Sec.~\ref{sec:HybridSimulation} we propose a hybrid simulation scheme, entailing a classical simulation of the main register and a quantum simulation of the ancilla register. For the sake of practical feasibility, in Sec.~\ref{sec:optimizing_graph_states} we also provide an optimization routine for the underlying graph states to minimize the number of required entangling gates. Finally, we show several applications demonstrating the utility of our algorithm in Sec.~\ref{sec:Applications}, more specifically the QAOA in Sec.~\ref{sec:qaoa} and the VQE in Sec.~\ref{sec:ucc}. 
Furthermore, we show some technical details in Appendices \ref{app:ParallelismAdapativity}--\ref{sec:data_vqe}.

\section{Preliminaries} \label{sec:MBQC}
To set the stage for our circuit-to-graph-state conversion algorithm, we first review some basic concepts needed to understand the idea.
We start by reviewing the definition of graph states and their connection to stabilizer states and then give a brief introduction to measurement-based quantum computing (MBQC).

\subsection{Graph- and stabilizer states} \label{sec:GraphStates}

The measurement-patterns obtained through our protocol are based on a computational resource state called graph state \cite{anders2006fast, hein2004multiparty}. 
An $N$-qubit graph state $\ket{G}$ is associated to an undirected graph $G=(V,E)$, whose $|V|=N$ vertices correspond to $N$ qubits prepared in the $\ket{+} \equiv 1/\sqrt{2}(\ket{0}+\ket{1})$ state, while the set of edges $E$ describes the action of controlled-$Z$ ($CZ$) operations between them. It can therefore be constructed as 
\begin{equation}
    \ket{G} = \left(\prod_{a,b\in E} CZ_{ab}\right)\left(\prod_{a\in V} H_a\right) \ket{0}^{\otimes N}.\label{eq:graph_state}
\end{equation}

\begin{figure}
    \centering
    \includegraphics[width=0.7\textwidth]{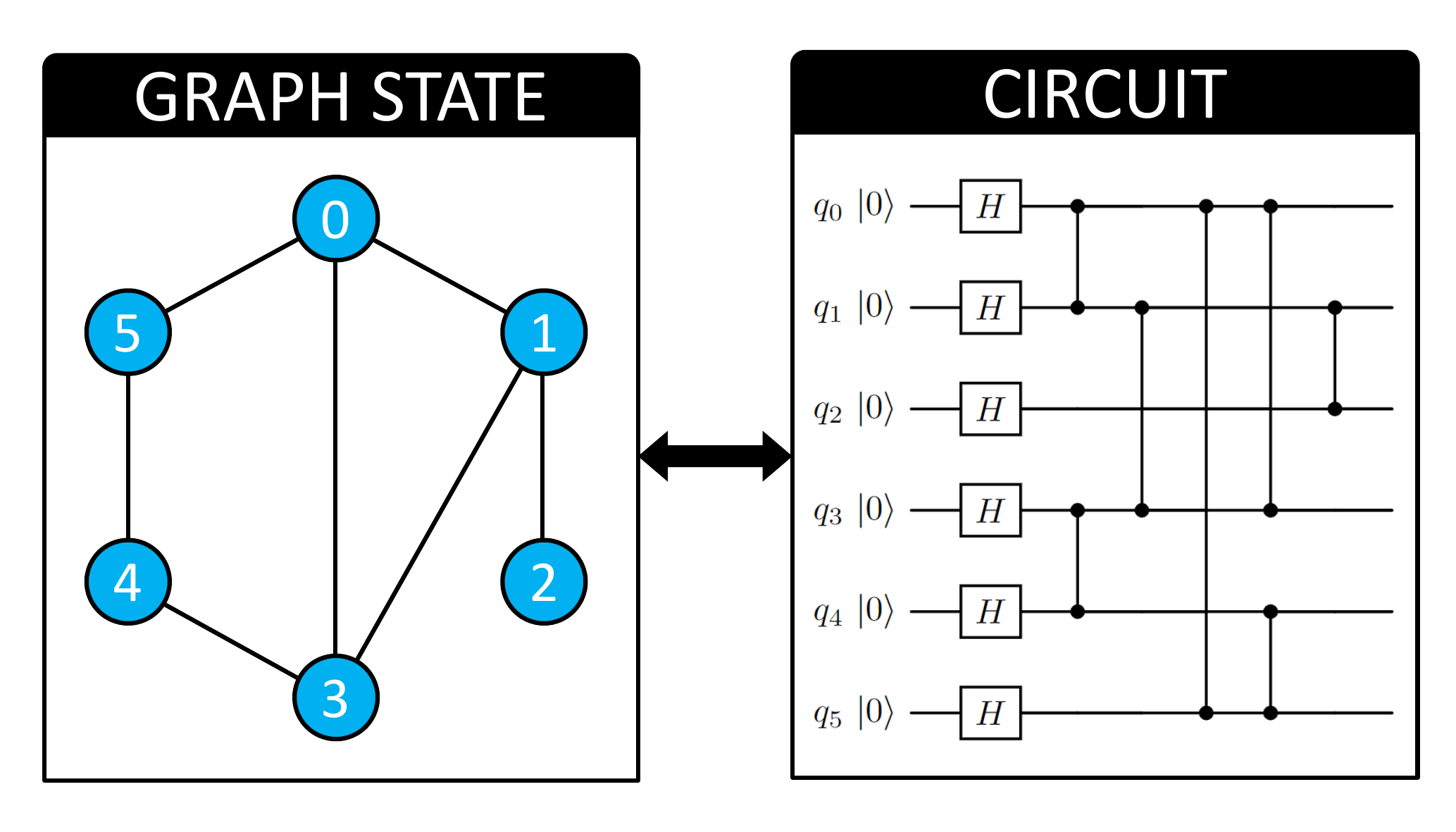}
    \caption{A graph state (left) and the corresponding quantum circuit (right). Some of the entangling gates can be applied in parallel, hence the total depth of the circuit in terms of entangling layers is only three.}
    \label{fig:GraphState}
\end{figure}

The action of all $CZ_{ab}$ gates commute with each other and can, in principle, be applied in parallel. 
In order to prepare a graph state on a quantum computer (using the circuit model), one thus needs a maximum depth of 
\begin{equation}
    d = \max_{\alpha\in V} |N_G(\alpha)|,
\end{equation}
where $N_G(\alpha)$ denotes the set of vertices connected to the vertex $\alpha$, i.e.,~$|N_G(\alpha)|$ is the degree of $\alpha$. An example of a graph state together with its corresponding quantum circuit is shown in Fig.~\ref{fig:GraphState}.

Eq.~\eqref{eq:graph_state} shows that any graph state on $N$ qubits can be generated through a sequence of Hadamard and $CZ$-gates. Both belong to the Clifford group $\mathcal{C}_N$, which is defined as
\begin{equation}
    \mathcal{C}_N = \{U \in \text{SU}(2^N) ~|~ U P U^\dagger \in \mathcal{P}_N ~\forall P \in \mathcal{P}_N\},
\end{equation}
where $\mathcal{P}_N$ denotes the $N$-qubit Pauli group. 
Any Clifford operator can be generated by three elementary gates (see e.g.~\cite{nielsen2002quantum}): the Hadamard gate ($H$), the phase gate ($S$) and the two-qubit $CZ$- or CNOT-gate. 

An $N$-qubit stabilizer state is a quantum state, that can be prepared by a sequence of Clifford gates acting on the $\ket{0}^{\otimes N}$ state.
Thus, by definition, every graph state is a stabilizer state -- the reverse is not true. 
However, one can show that every stabilizer state is local Clifford equivalent to a graph state (LC-equivalence), i.e.,~for every stabilizer state a graph state can be found, that can be transformed to the stabilizer state using one-qubit Clifford gates only~\cite{van2004graphical, schlingemann2001stabilizer}.

Quantum circuits consisting only of Clifford gates can be simulated efficiently on a classical computer according to the Gottesman-Knill theorem~\cite{gottesman1998heisenberg}.
Using the LC-equivalence of stabilizer states with graph states, one can simulate an $N$-qubit Clifford circuit using $\mathcal{O}(N \ln N)$ space in computer memory.
The core idea is to store the graph together with the local Clifford operations (also called vertex operators or VOPs) for each qubit \cite{anders2006fast}.

\subsection{Measurement-based quantum computing} \label{sec:MBQC2}
In the model of measurement-based quantum computing (MBQC), quantum computations start from a highly entangled many-qubit state, called resource state, which is modified by applying a sequence of adaptive measurements onto a subset of qubits. 
At first sight, it might seem counter-intuitive that universal quantum computation can be performed using irreversible, destructive measurements. 
While MBQC involves a loss of information concerning the entire resource state, it still performs unitary transformations on the subset of qubits that are not measured during the computation.
On these unmeasured qubits any unitary operation can be implemented, provided the resource state is sufficiently complex.

Although different MBQC schemes exist, here we focus on the so-called cluster model or One-Way Quantum Computer (1WQC) of Raussendorf and Briegel \cite{raussendorf2001one, raussendorf2001computational, , PhysRevA.68.022312, raussendorf2009measurement}. 
A review of 1WQC and other measurement-based schemes can be found in Ref.~\cite{jozsa2006introduction}. 

In MBQC, all quantum gates are implemented as a sequence of single-qubit measurements on a suitable large cluster state. 
The measurement basis needed for universal quantum computing is given by
\begin{equation}
M(\theta) = \{\ket{0} \pm e^{i\theta}\ket{1}\}.
\end{equation}
The measurement in the $M(\theta)$-basis is achieved by applying the unitary $H R_z(\theta)$ to the computational basis and performing the usual $z$-measurement.

\section{Mapping circuits to graph states} \label{sec:MappingGraphStates}

In this section, we introduce an algorithm, that allows the mapping of a quantum circuit to a graph state, that can then be used within the MBQC protocol. The core idea is to map unitaries of the form
\begin{equation}
    U = e^{-\frac{i}{2}\theta \mathcal{P}},
    \label{eq:pauli_unitary}
\end{equation}
where $\mathcal{P}$ denotes an $N$-qubit Pauli string $\mathcal{P} \in \{I, X, Y, Z\}^{\otimes N}$, to an ancilla qubit in the circuit.
To implement such operators in the circuit model, the Pauli string is diagonalized by applying local Clifford operators using the identities $X=HZH$ and $Y=S H Z H S^\dagger$. 
This effectively reduces the operator pool to $\mathcal{P}'\in \{I, Z\}^{\otimes N}$. 

\begin{figure}
    \centering
    \includegraphics[width=0.95\textwidth]{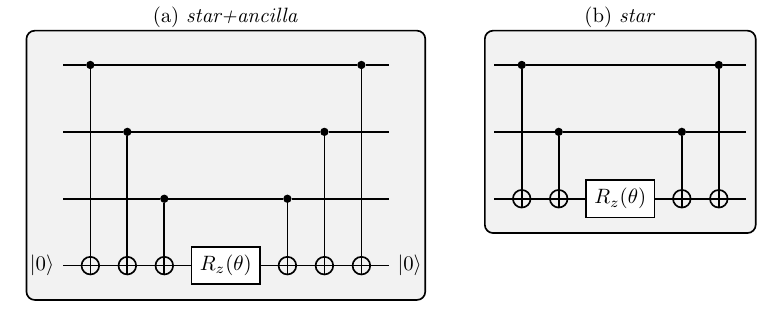}
    \caption{Implementations of the operator $\exp{(-i/2\theta Z_0Z_1Z_2)}$ using (a) the \textit{star}+\textit{ancilla} layout, and (b) the pure \textit{star} layout. The term \textit{star} refers to the star-shaped structure of the entangling gates.}
    \label{fig:SSA}
    \phantomlabel{a}{fig:SA}
    \phantomlabel{b}{fig:S}
\end{figure}

We review two common circuit structures representing $\exp{(-i/2\theta \mathcal{P}')}$. 
In the \textit{star+ancilla} layout (Fig.~\ref{fig:SA}), all non-identity qubits of the string are directly entangled with an ancilla qubit (star-like structure), which is initialized in state $\ket{0}$ and on which the $R_z$-rotation is carried out. 
By repeating the same entanglement structure, the entanglement with the ancilla qubit is undone. 
The star-like entanglement with the ancilla can be interpreted as a computation of the parity of the $N$ qubits in a classical manner. 
For an even parity, a phase shift of $\exp{(-i\theta/2)}$ is applied, otherwise it is $\exp{(i\theta/2)}$. 
Finally, the parity is uncomputed, erasing the ancilla and leaving it in the $\ket{0}$ state again~\cite{nielsen2002quantum}.

The second structure, which we refer to as the \textit{star} layout (Fig.~\ref{fig:S}), works similarly, with the key difference that the entanglement and $R_z$-gate is performed with respect to one of the non-identity qubits instead of an ancilla. Both circuits are equivalent and have their own benefits depending on the problem at hand. 
Further, it should be mentioned that both circuits can be equivalently realized using a ladder-like entanglement structure, though this approach is not further discussed within our work due to less convenient gate cancellation properties \cite{gui2020term}.

\subsection{Replacing single-qubit gates with measurements} \label{sec:Replace}

\begin{figure}[b!]
    \centering
    \includegraphics[width=0.49\textwidth]{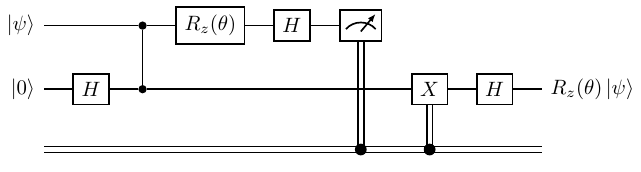}
    \hfill
    \includegraphics[width=0.49\textwidth]{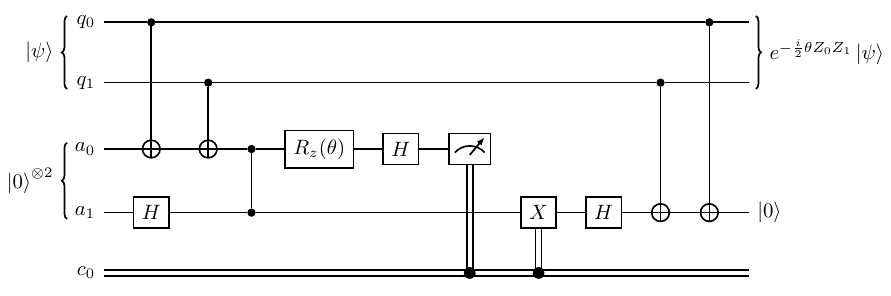}
    \caption{Left panel: Measurement-pattern to implement the $R_z$-gate. After measurement of the first qubit, the second qubit is in state $X^{s} H R_z(\theta)\ket{\psi}$. Right panel: Circuit implementation of the unitary $\exp{(-i/2\theta Z_0 Z_1)}$ using the \textit{star+ancilla} layout, after insertion of the pattern. The ancilla qubit $a_1$ is in the $\ket{0}$-state at the beginning and at the end.}
    \label{fig:MBRZ}
\end{figure}

The first ingredient in converting a given quantum circuit into a graph state is to replace all single-qubit rotations $R_z(\theta)$ by a measurement pattern by introducing one ancilla qubit.
Consider the example of a single qubit in an arbitrary state $\ket{\psi}$ entangled with a second qubit in the $\ket{+}$ state via a $CZ$-gate. 
After measuring the first qubit in the $M(\theta)$ basis, the second qubit is left in the state 
\begin{equation}
\ket{\psi}^\prime = X^{s} H R_z(\theta)\ket{\psi},
\end{equation}
where $s\in \{0, 1\}$ is the measurement outcome on the first qubit. 
Acting with $X^{s}$ and then with $H$ on the second qubit yields the deterministic final state $R_z(\theta)\ket{\psi}$, which is the desired $R_z(\theta)$-gate.
The quantum circuit performing this operation is shown in Fig.~\ref{fig:MBRZ} and can be used as a pattern to replace any $R_z$-gate in a given circuit.

Next, we use this pattern to rewrite unitaries defined by Eq.~\eqref{eq:pauli_unitary} in the MBQC protocol.
As an example, let us consider the unitary $\exp{(-i /2\theta Z_0 Z_1)}$. 
We use the \textit{star+ancilla} layout as starting point to exemplify some aspects of our algorithm. 
It is straightforward to derive the pattern for the same example in the \textit{star} layout.
By replacing the $R_z$-gate on the ancilla qubit with the pattern, one derives the circuit shown in the right panel of Fig.~\ref{fig:MBRZ}.

\begin{figure}
    \centering
    \includegraphics[width=0.7\textwidth]{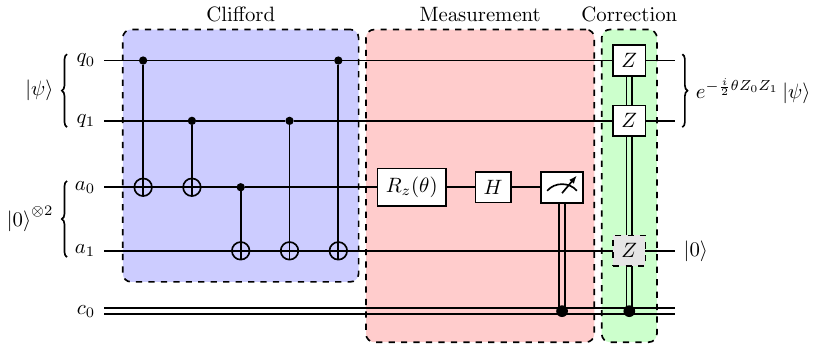}
    \caption{Pattern to implement $\exp{(-i/2\theta Z_0 Z_1)}$ acting on qubits $q_0$ and $q_1$, after shifting the Pauli corrections to the end. The final $Z$ correction on the ancilla $a_1$ (in gray) can be left out since it does not affect the zero-state.}
    \label{fig:ZZStarAncillaMBQC2}
\end{figure}

So far, it appears that there is no benefit from replacing the $R_z$-gate with the MB-protocol. 
It is rather the opposite: One more $CZ$-gate and an additional measurement are required to realize the same operation.
However, by shifting the classically-controlled Pauli corrections across the Clifford gates to the end of the circuit, the quantum circuit can be separated into three components: a pure Clifford layer, a measurement layer and a correction layer, as shown in Fig.~\ref{fig:ZZStarAncillaMBQC2}.
In accordance with Ref.~\cite{BROADBENT20092489} we call this the standard form of a pattern.

The standard form of a pattern can be easily achieved by employing the following identities for single-qubit
\begin{align}
    H X &= Z H, \qquad H Z = X H, \nonumber\\
    S^{(\dagger)} X &= Y S^{(\dagger)},\qquad
    S^{(\dagger)} Y = X S^{(\dagger)},
    \label{eq:ShiftingRules1}
\end{align}
and two-qubit Clifford gates
\begin{align}
    CX_{12}(I_1\otimes Z_2) &= (Z_1\otimes Z_2) CX_{12} \nonumber\\
    CX_{12}(X_1\otimes I_2) &= (X_1\otimes X_2) CX_{12} \nonumber\\
    CZ_{12}(I_1\otimes X_2) &= (Z_1\otimes X_2) CZ_{12}.
\label{eq:ShiftingRules2}
\end{align}
All remaining identities can be directly obtained from $Y \propto Z\cdot X$ and $CZ_{12}=CZ_{21}$. 

Interestingly, the ancilla qubit $a_1$ in the pattern shown in Fig.~\ref{fig:ZZStarAncillaMBQC2} is always left in the $\ket{0}$-state, regardless of the intermediate measurement outcome. 
This is a general feature for a pattern in the \textit{star+ancilla} layout.
The last classically-controlled $Z$ gate on the last ancilla can thus always be neglected, since $Z\ket{0} = \ket{0}$. 

\subsection{Quantum parallelism and adaptive measurements} \label{sec:Parallelism}

\begin{figure}
    \centering
    \includegraphics[width=0.95\textwidth]{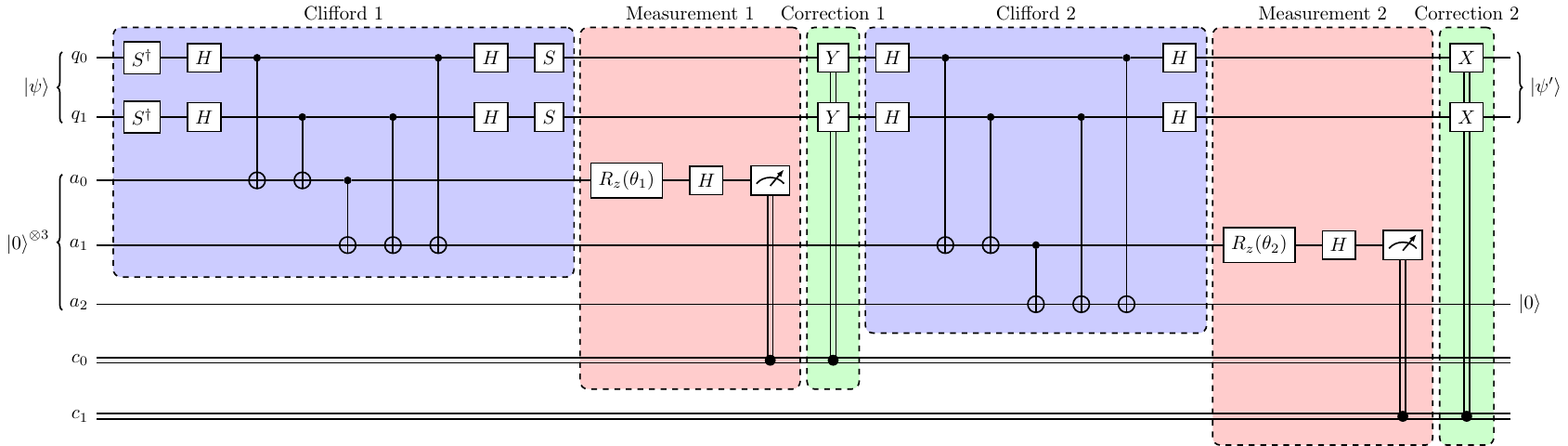}
    \hfill
    \par\bigskip
    \includegraphics[width=0.95\textwidth]{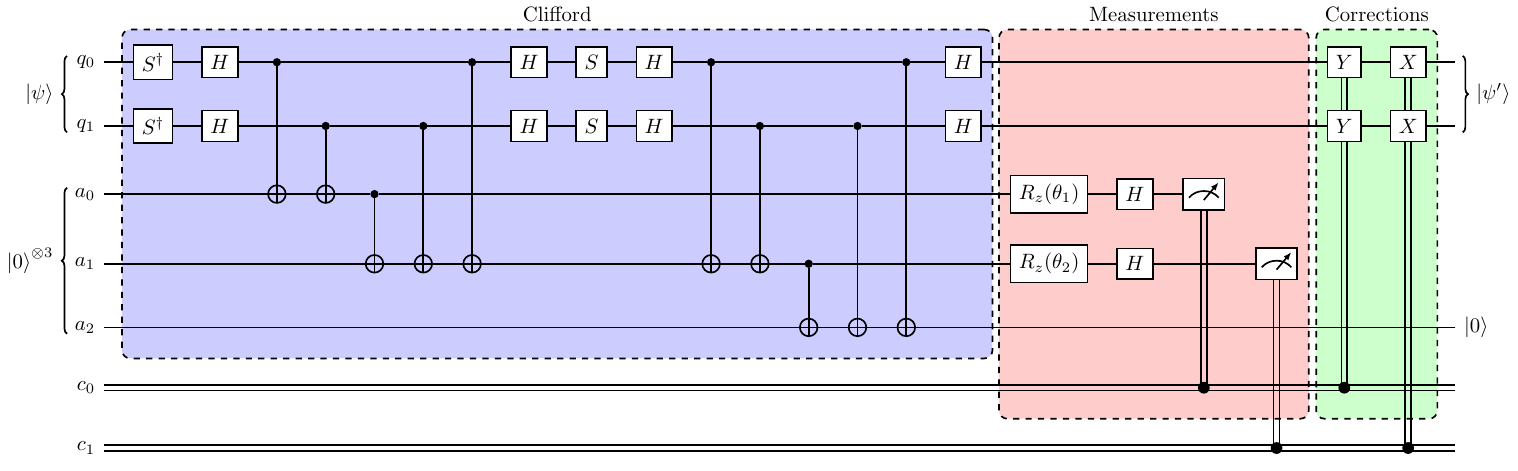}
    \caption{Upper panel: Naive circuit representation obtained through concatenation of the circuits for $\exp{(-i/2\theta_1Y_0 Y_1)}$ and $\exp{(-i/2\theta_2 X_0 X_1)}$. Lower panel: Measurement-based pattern to implement $\exp{[-i/2(\theta_2 X_0 X_1 + \theta_1Y_0 Y_1)]}$ acting on the first two qubits. By rewriting the pattern into its standard form, the measurements of the two ancilla can be applied in parallel.\label{fig:XXZZ}}
\end{figure}

\begin{figure}
    \centering
    \includegraphics[width=0.95\textwidth]{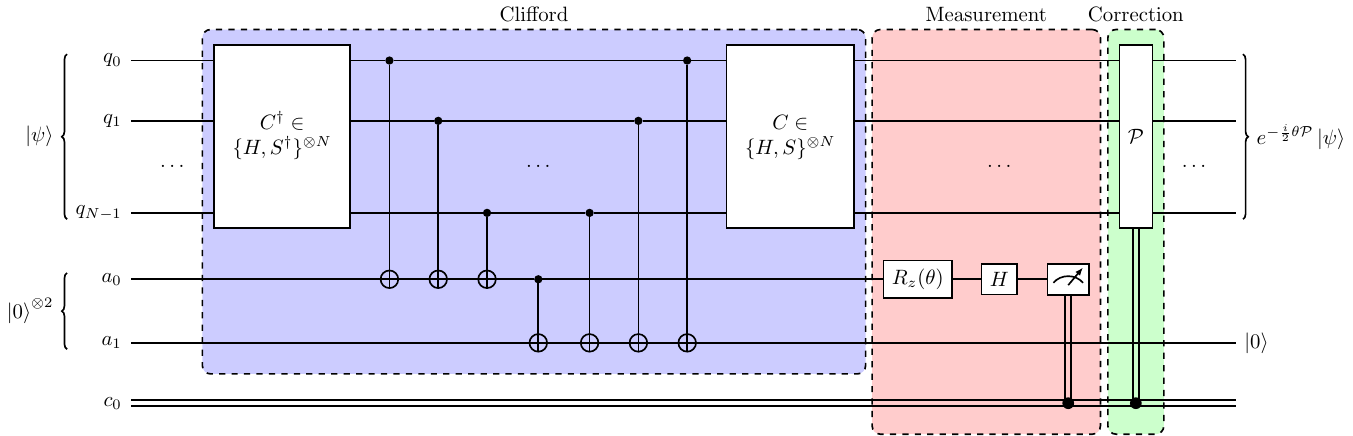}
    \hfill
    \par\bigskip
    \includegraphics[width=0.95\textwidth]{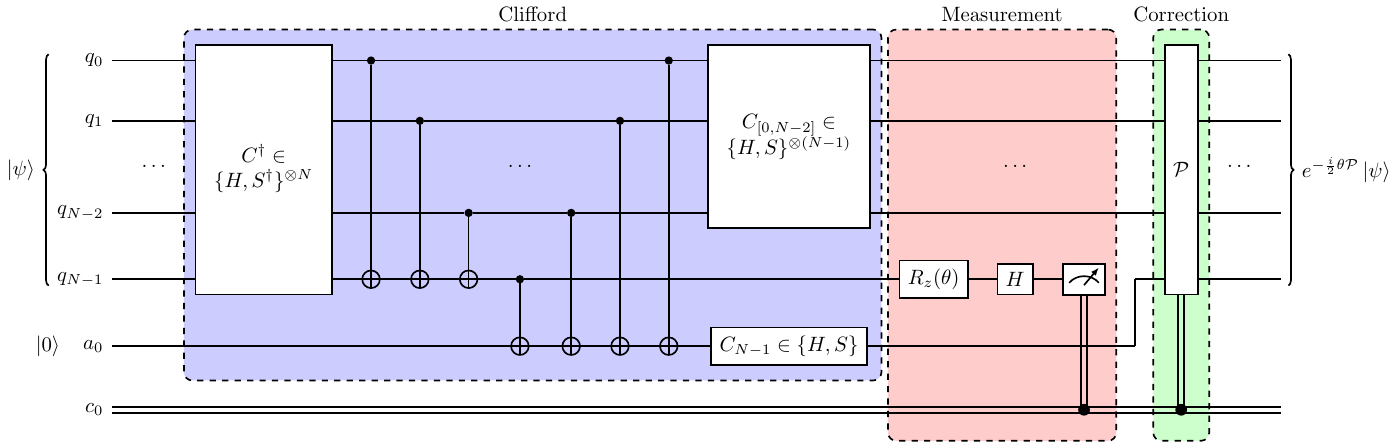}
    \caption{Patterns to implement $\exp(-i/2\theta \mathcal{P})$ with the \textit{star+ancilla} layout (top panel) and the \textit{star} layout (bottom panel). Here, we assume that $\mathcal{P}$ involves no identities, thus the Clifford circuit entangles all main qubits with the ancilla register. The final correction is given by the Pauli string $\mathcal{P}$ itself. Note that in the bottom panel, the ancilla qubit $a_0$ takes the role of the last main qubit $q_{N-1}$ after the measurement.}
    \label{fig:PauliStringCircuit}
\end{figure}

To showcase an important property of our MB-circuit protocol, we now consider the case of two commuting Pauli strings $Y_0 Y_1$ and $X_0 X_1$, once again implemented through the \textit{star+ancilla} layout. 
The naive circuit representation, which is obtained through concatenation of the circuits for $\exp{(-i/2 \theta_2 X_0 X_1)}$ and $\exp{(-i/2\theta_1 Y_0 Y_1)}$, is depicted in the upper panel of Fig.~\ref{fig:XXZZ}. 

Using the same circuit identities as before, we can shift the first correction and measurement layer across the second Clifford layer. 
In this particular case, this step introduces no corrections on the second ancilla qubit, which would have to be carried out before the second measurement. 
Instead, both measurements can be performed in parallel (see lower panel of Fig.~\ref{fig:XXZZ}).
The fact, that the measurements can be parallelized here, is no coincidence. In the following, we derive the condition for parallelism. 

We already know, that for the application of a generic Pauli exponential, Pauli corrections are applied to all qubits corresponding to non-identity operations of the Pauli string (e.g.,~if the Pauli string inside the exponential reads $X_1 Z_3 Y_6$, Pauli corrections are applied to qubits $1$, $3$ and $6$ only). From Fig.~\ref{fig:ZZStarAncillaMBQC2} and Eq.~\eqref{eq:ShiftingRules1}, we can infer that the correction is precisely given by a controlled version of the Pauli string itself, since the $Z$-corrections undergo a basis transformation according to the string. Schematically, this is shown in Fig.~\ref{fig:PauliStringCircuit} for both layouts (\textit{star} and \textit{star+ancilla}).

Let us assume, we want to apply two unitaries generated by the Pauli strings $\mathcal{P}$ and $\tilde{\mathcal{P}}$.
We now investigate the conditions, which these two Pauli strings have to fulfill, for parallel measurement. 
As explained before, the circuit implementing the matrix exponential of $\mathcal{P}$ ends with a controlled version of $\mathcal{P}$ itself, i.e., $\mathcal{P}^s$, where $s\in \{0,1\}$. Hence, if we want to bring the pattern implementing the product $\exp(-i/2\tilde\theta\tilde{\mathcal{P}})\exp(-i/2\theta\mathcal{P})$ into a standard form, it is sufficient to see what happens when shifting $\mathcal{P}^s$ through $\exp(-i/2\tilde\theta\tilde{\mathcal{P}})$. 

In Appendix~\ref{app:ParallelismAdapativity} we show that:

\begin{equation}
    \label{eq:correction_commutation}
    e^{-\frac{i}{2}\tilde\theta\tilde{\mathcal{P}}} \mathcal{P}^s = \mathcal{P}^s
    \begin{cases}
        e^{-\frac{i}{2}\tilde\theta\tilde{\mathcal{P}}}, & \text{if $[\mathcal{P}, \tilde{\mathcal{P}}]=0$} \\
        e^{-\frac{i}{2}(-1)^s\tilde\theta\tilde{\mathcal{P}}}, & \text{else}
    \end{cases}\ ,
\end{equation}
where $s$ denotes the measurement outcome of the first ancilla qubit.
Consequently, the product of two unitaries generated by two Pauli strings can be implemented in parallel only if the strings commute.
Otherwise, the rotation angle $\tilde \theta$ of the second Pauli exponential has to be adapted to the measurement outcome $s$ of the first ancilla, leading to an adaptive, i.e., non-parallel measurement pattern.

Generalizing this result to the application of $M$ unitatries generated by Pauli strings $\{\mathcal{P}_1, \mathcal{P}_2, \dots, \mathcal{P}_M\}$, we find that the final correction of the pattern implementing this operation in standard form is given by
\begin{equation}
    \prod_{m=1}^M \mathcal{P}_m^{s_m},
    \label{eq:final_correction}
\end{equation}
where $s_m$ is the measurement outcome of the $m$-th ancilla. Eq.~\eqref{eq:final_correction} allows to write down the final correction of an arbitrary measurement pattern, without the additional computational cost of propagating all corrections to the end of the circuit. 

All ancilla qubits can be measured in parallel, only if all Pauli strings commute with each other.
Otherwise, the measurement bases have to be adapted according to Eq.~\eqref{eq:correction_commutation}. 
In contrast to the final correction in Eq.~\eqref{eq:final_correction}, the adaptive measurement bases depend on the order in which the Pauli exponentials are implemented. 
To implement the unitary $\exp{(-i/2 \theta_i \mathcal{P}_i)}$, the rotation angle of the measurement basis is obtained by flipping the sign of $\theta_i$ for each previous non-commuting pattern measured in the $\ket{1}$-state, i.e.,
\begin{equation}
     \theta_i \to (-1)^{h_i} \theta_i, \quad \text{where} \quad  h_i = \sum_{\substack{j < i\\ [\mathcal{P}_i, \mathcal{P}_j] \neq 0}} s_j.
     \label{eq:adaptive_basis}
\end{equation}
To minimize the number of adaptive measurements, it is therefore convenient to first sort the generating strings into fully commuting groups.

\subsection{Applying quantum gates in constant depth in the circuit model} \label{sec:ParallelClifford}

\begin{figure}
    \centering
    \includegraphics[width=0.49\textwidth]{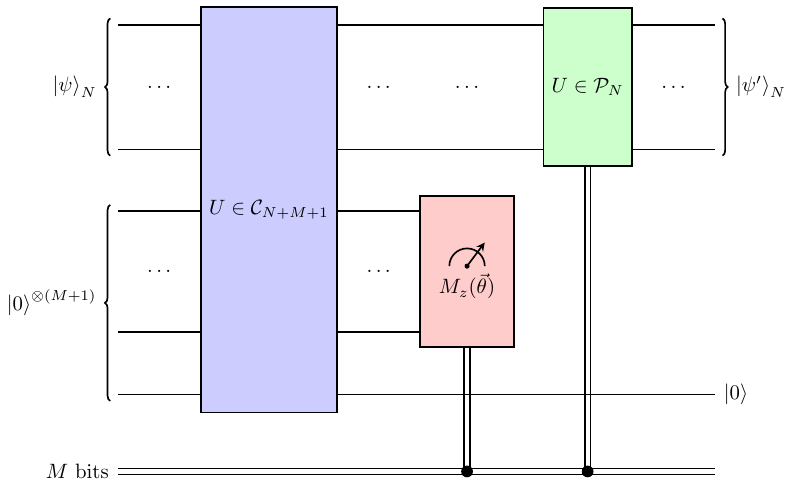}
    \hfill
    \includegraphics[width=0.49\textwidth]{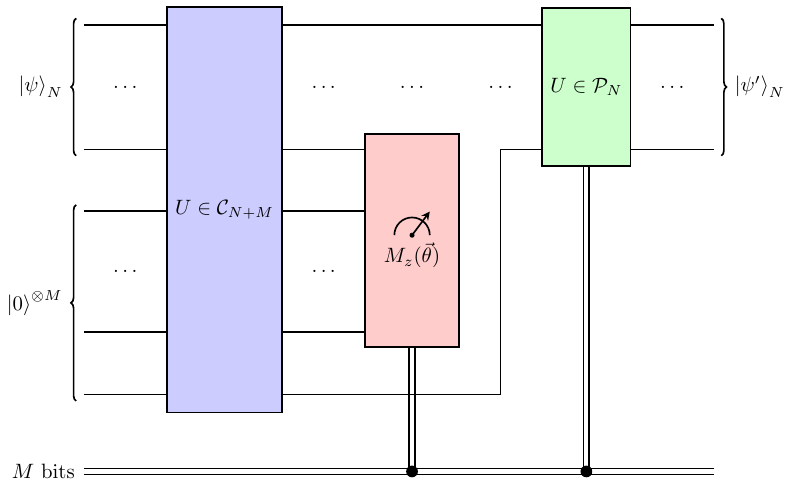}
    \caption{General circuit structure for $M$ Pauli strings acting on $N$ qubits using MBQC on arbitrary initial states with different architectures for the multi-qubit rotations. The left panel depicts the structure obtained through the \textit{star+ancilla} layout (Fig.~\ref{fig:SA}), therefore the last ancilla is always left in state $\ket{0}$. The right panel shows the circuit using the \textit{star} layout (Fig.~\ref{fig:S}), hence the first measurement occurs on the main register, thus teleporting the last main qubit through the entire ancilla register.}
    \label{fig:BuildingBlocks}
\end{figure}

In this section, we show how our method can be used to derive measurement patterns, that can be used to apply several commuting operators in parallel with constant circuit depth in the circuit model.
In the following, we assume that the depth of a quantum circuit is defined by the number of layers with at least one CNOT-gate, which are needed to implement it.

Let us assume that we apply several commuting unitaries $U_1,\dotsc, U_n$ generated by $n$ Pauli strings to a quantum state $\ket{\psi}_N$ -- the order does not matter, since they commute.
Then, we first derive the measurement pattern which implements $U=U_n \cdots U_2 \cdot U_1$ using the method outlined in the previous section.
This pattern, in standard form, has three layers: Cliffords, measurements (in parallel) and corrections, c.f.~Fig.~\ref{fig:BuildingBlocks}.
Since the measurement and correction layer have already constant depth, we just need to implement the Clifford operations in constant depth.

\begin{figure}
    \centering
    \includegraphics[width=0.98\textwidth]{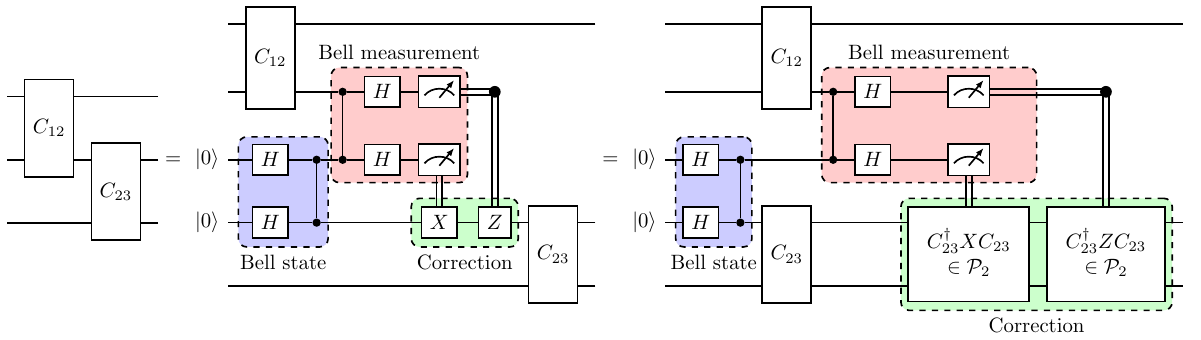}
    \caption{Parallelization of two-qubit Clifford gates by introducing ancilla qubits. By using quantum teleportation, any Clifford circuit can be reduced to constant depth.}
    \label{fig:ParallelClifford}
\end{figure}

For this, note that a general Clifford circuit can always be expressed as a sequence of one-qubit Cliffords and CNOT-gates.
Thus, it suffices to show that two sequential CNOT-gates (with potentially intermediate one-qubit Cliffords) can be applied in parallel in constant depth.
This can be achieved using the quantum teleportation algorithm.
The general construction is depicted in Fig.~\ref{fig:ParallelClifford}: Any sequence of two-qubit Clifford gates can be recast to a quantum process of constant depth.
Using this technique, the number of ancilla qubits grows linearly with the depth of the Clifford layer, while the additional classical computation due to the corrections grows logarithmically \cite{jozsa2006introduction}.

To summarize, we can implement arbitrary Clifford circuits as constant depth circuits. 
Combining this with the previous result of deriving a pattern for a group of commuting operators (Sec.~\ref{sec:Parallelism}), we conclude that our algorithm allows the implementation of several commuting operators as a constant-depth measurement-based pattern. 
More precisely, these constant-depth patterns can always be achieved with three entangling layers (one for the Bell state preparation of the ancilla qubits, one for the initial entangling gates and one for the Bell basis measurements), a measurement layer, and a corrective Pauli layer.

\subsection{Simulation and correction of the main qubits} \label{sec:HybridSimulation}

\begin{figure}
    \centering
    \includegraphics[width=0.95\textwidth]{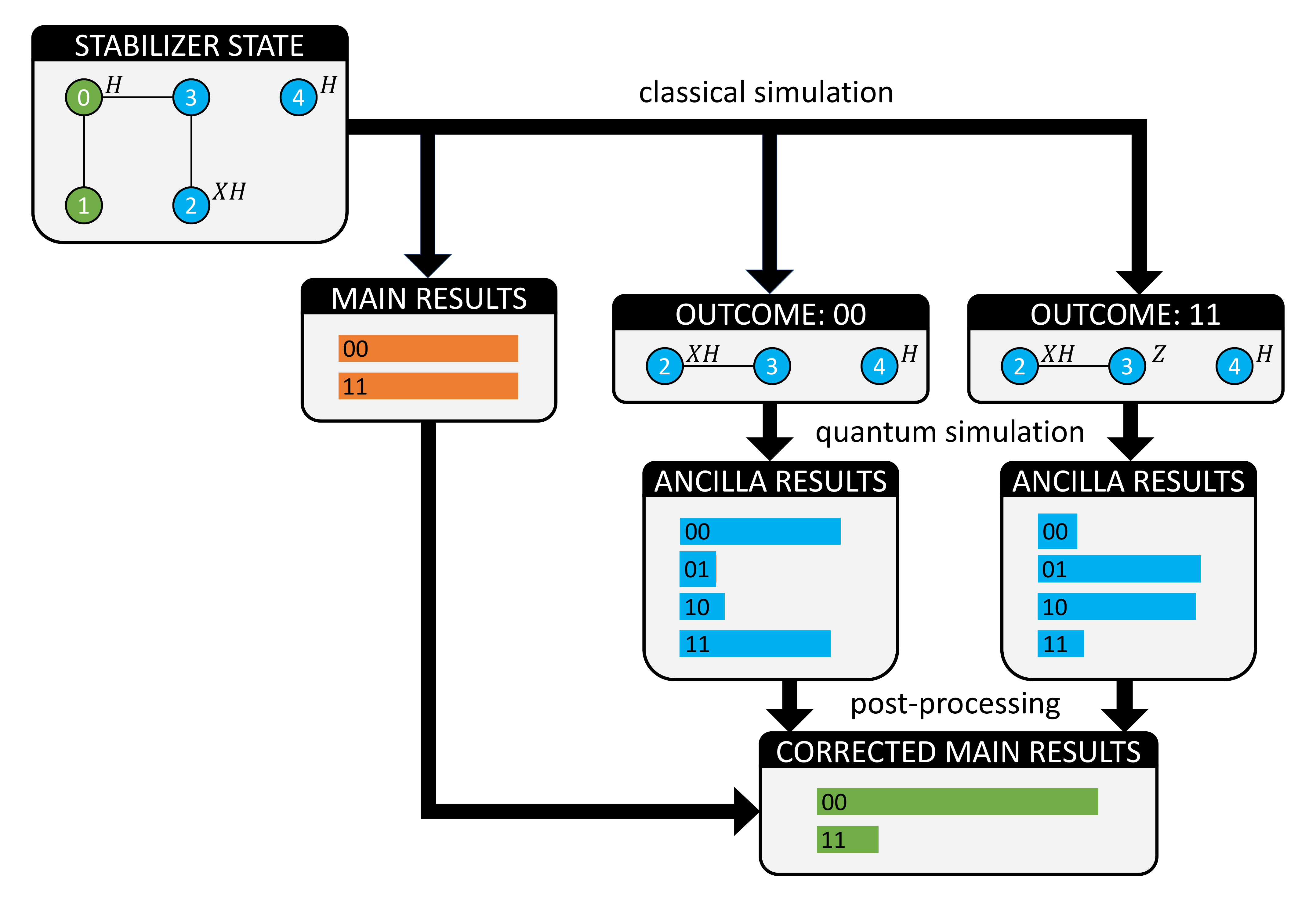}
    \caption{Overview how to simulate the main qubits. Starting with a graph state, in which the main qubits are measured in Pauli bases and the ancilla qubits in rotated bases, we can classically simulate the outcome of the main qubits. For each outcome, we derive the remaining state of the ancilla qubits before the measurement, which are all equivalent up to local unitaries.}
    \label{fig:HybridSim}
\end{figure}

Up to this point, our observations hold for arbitrary input states.
We now show how the circuits can be further reduced by simulating the Clifford part classically, assuming that the main qubits are initialized in a stabilizer state, i.e.,~the initial state can be prepared with Clifford gates acting on the $\ket{0}^{\otimes N}$ state.

We start by converting the Clifford part of the circuit into a graph state using the Clifford simulation algorithm outlined in Ref.~\cite{anders2006fast} (the original code from the authors can be found online\footnote{\url{https://github.com/marcusps/GraphSim}}), which we implemented with the \texttt{rustworkx} package~\cite{Treinish2022}.
The result of this conversion is a graph state of size $N_q+N_a$, where $N_q$ denotes the number of initial qubits (main qubits) and $N_a$ the number of required ancilla qubits, which is equivalent to the number of non-Clifford gates in the initial circuit.
In this graph state, all main qubits are measured in a Pauli basis, while the ancilla qubits are measured in the $M(\theta)$ basis.
Furthermore, after measuring the ancilla qubits, a final Pauli correction has to be applied to the main qubits.

In the MBQC protocol, one would now proceed by first measuring the ancilla qubits, correcting the main qubits depending on their outcome and then, at the end, perform measurements on the main qubits depending on the observables one wishes to extract. 
However, if the main qubits are measured in the Pauli $X$, $Y$ or $Z$-basis, we can equally first simulate the measurement outcome of the main qubits.
This might seem surprising, but it can be explained as follows.

Suppose we have a main qubit $q$, which is corrected by Pauli $X$ depending on the measurement-outcome $s_a$ of an ancilla qubit $a$.
Then, after the measurement of the ancilla qubit, the new state of $q$ is given by $X^{s_a} \ket{q}$.
If we now measure $q$, we know that, depending on $s_a$, the state $\ket{q}$ was either $\ket{0}$ or $\ket{1}$ \textit{before} the measurement of $\ket{a}$.
Thus, we can equivalently flip the measurement outcome $s_q$ of $\ket{q}$ depending on $s_a$.
In the case of a $Z$-correction nothing has to be done. 
For a $Y$-correction we can always rewrite $Y=X\cdot Z$ up to an irrelevant, global phase.

Following this logic, we can first efficiently simulate the measurements of all main qubits using the graph state simulator, then execute the remaining circuits on the ancilla qubits and use a post-processing algorithm to correct the counts of the main qubits accordingly.
More specifically, we first classically sample one shot on the main qubits neglecting the correction layer. 
Based on that, we obtain a bit string and a result-dependent stabilizer state (which is equivalent up to local unitaries to a graph state) for the ancilla qubits. 
This stabilizer state is then prepared and measured in the rotated bases on a quantum computer. 
If the ancilla measurement results imply an $X$ or $Y$ correction on the main qubits, the bit string is modified through the appropriate bit
flips. 

It is important to emphasize, that, while an exponential amount of classical measurement outcomes can occur, there is a one-to-one correspondence between a classically simulated shot in our algorithm and a quantum shot in the circuit model. 
Hence, the number of classical results that actually have to be considered is limited by the number of shots. 
Consequently, no exponential blowup occurs in this hybrid approach.

As an example, let us reconsider the measurement pattern that implements $\exp{[-i/2\theta(X_0X_1+Y_0Y_1)]}$ from Fig.~\ref{fig:XXZZ} and let it act on the initial state $\ket{00}$.
The overview over the full calculation is summarized in Fig.~\ref{fig:HybridSim}.
After simulating the Clifford part of the circuit, we find the graph state shown in the top left.
By using the graph-based sampling algorithm, we find two possible measurement-outcomes for the two main qubits: $00$ and $11$.
Since we sample from a stabilizer state, all bit-strings (with non-zero amplitudes) appear with equal probability~\cite{dehaene2003clifford}.
All outcomes lead to local equivalent stabilizer states, where the ancilla qubits are decoupled from the main qubits.
In our example, we find the two stabilizer states shown in the second row of Fig.~\ref{fig:HybridSim}.
Running these stabilizer state on a quantum computer results in the probability distribution that is needed to correct the counts of the main qubits.

\subsection{Optimization of graph states}
\label{sec:optimizing_graph_states}

\begin{figure}[b!]
    \centering
    \includegraphics[width=\textwidth]{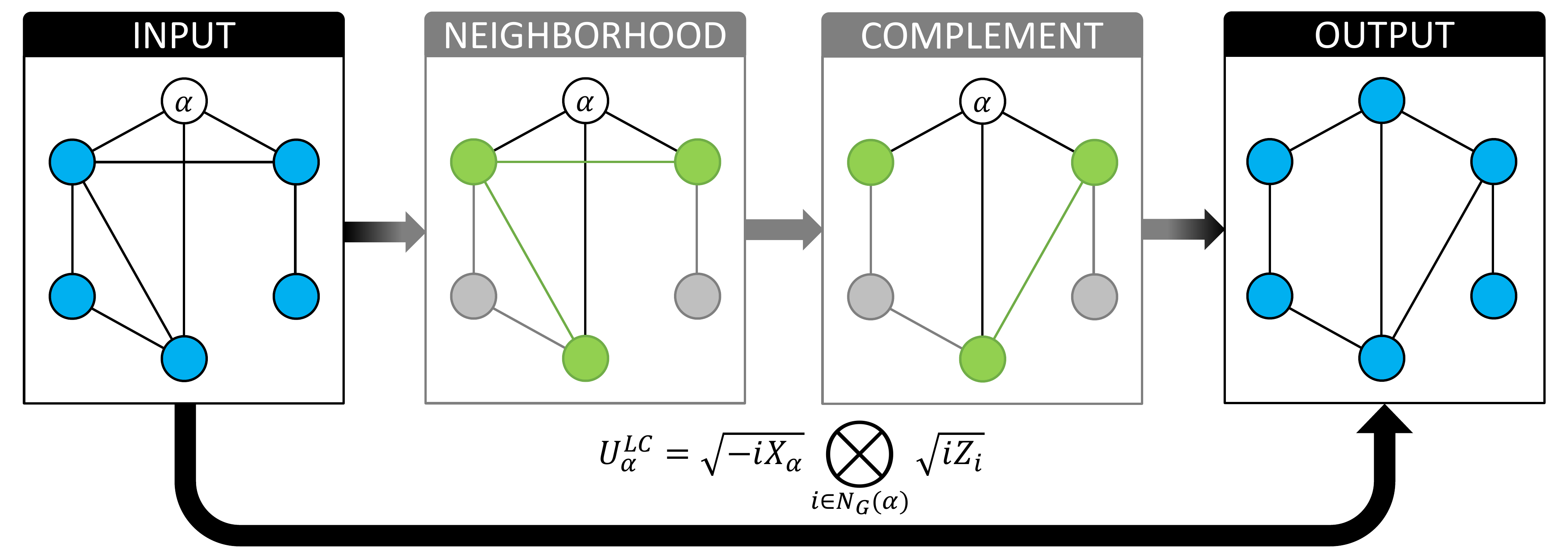}
    \caption{Local complementation of vertex $\alpha$. In this example, the number of edges (i.e.~the number of $CZ$ gates to prepare the graph state) is reduced by one.}
    \label{fig:LC}
\end{figure}

The graph states obtained through our simulation protocol are often quite complicated.
However, they are not unique.
Stabilizer states are invariant under the operation of local complementations.
In the following, we denote the local complementation of a graph at a vertex $\alpha$ by $\text{LC}_\alpha$.
Applying $\text{LC}_\alpha$ to a graph $G(V, E)$ complements the neighborhood of the vertex $\alpha$. 
That is, all existing edges between the vertices in $N_G(\alpha)$ are removed and all missing edges in $N_G(\alpha)$ are added. 
The corresponding stabilizer state is preserved by applying the local unitaries \cite{anders2006fast, adcock2020mapping}:
\begin{equation}
    U_\alpha^{\text{LC}} = \sqrt{-i X_\alpha}\bigotimes_{i\in N_G(\alpha)} \sqrt{iZ_i},
\end{equation}
where $N_G(\alpha)$ denotes the neighborhood of the vertex $\alpha$.
The procedure is depicted in Fig.~\ref{fig:LC}. Note that the total number of edges may be changed after the operation.

With current NISQ-hardware in mind, we want to find optimal graph states with respect to the number of edges, which defines the number of $CZ$ gates in the preparation circuit. Alternatively, one could optimize the states with respect to the maximum degree, thus minimizing the circuit depth required for preparation, or even optimize a trade-off between both properties.  

To perform the optimization task, we employ the simulated annealing algorithm \cite{van1987simulated}, which we will briefly outline here. A more detailed description is provided in Appendix \ref{app:SimulatedAnnealing}. The solution space is the set of graph states that are LC-equivalent to the initial graph state obtained by converting the Clifford circuit. 
In each iteration, a random node of the graph is locally complemented. 
The cost function we aim to minimize is then evaluated with respect to the new graph state. 
If it is improved, the old graph state is discarded. Otherwise, we might still keep the new graph state, but only with steadily decreasing probability according to a Boltzmann distribution.

Despite its inherent simplicity and no guarantee to find the global optimum, we have observed major reductions in circuit depth results using this method.

\section{Applications} \label{sec:Applications}
In this section, we show how our methods can be applied directly to two important NISQ algorithms.
In Sec.~\ref{sec:qaoa} we show how combinatorial problems can be solved using the QAOA and in Sec.~\ref{sec:ucc} we show how the electronic-structure problem can be tackled by using the VQE. All simulations were performed using \texttt{qiskit}~\cite{Qiskit}.

For the QAOA we derive an ansatz containing mid circuit measurements due to non-commuting operators.
For the VQE we design an ansatz which consists entirely of a set of commuting operators.
In this scenario our method yields the most powerful reduction in circuit depth and does not rely on mid-circuit measurements, which are still challenging on current quantum computing devices.

\subsection{QAOA and max-cut problems\label{sec:qaoa}}
The QAOA \cite{farhi2014quantum, zhou2020quantum} is an optimization algorithm designed to solve combinatorial optimization problems in the NISQ-era \cite{preskill2018quantum}.
The idea is to encode the optimization problem into a minimization problem of a generic Ising-Hamiltonian (also called cost Hamiltonian $H_c$)
\begin{equation}
    H_c = \sum_{i<j} w_{ij} Z_i Z_j+\sum_i b_i Z_i,
\end{equation}
where $w_{ij}$ and $b_i$ are coefficients depending on the optimization problem.
In its standard formulation, the QAOA algorithm tries to find the lowest energy (corresponding to the optimal solution of the initial problem) using a variational ansatz of depth $p$, which is given by
\begin{equation}
    \ket{\psi} = e^{-i\beta_p/2 H_m }e^{-i\gamma_p/2 H_c }\cdots e^{-i\beta_1/2 H_m }e^{-i\gamma_1/2 H_c } \ket{+}^{\otimes N},
\end{equation}
where  $H_m \equiv \sum_{i=1}^{N} X_i$ is the so-called mixer Hamiltonian, and $N$ denotes the number of qubits.
The $\gamma_i$ and $\beta_i$ are in total $2\,p$ variational parameter, which are obtained in a classical optimization feedback and aim to minimize the expectation value $\braket{\psi | H_c | \psi}$, which can be estimated efficiently on a quantum computer.

\begin{figure}
    \centering
    \includegraphics[width=0.25\textwidth]{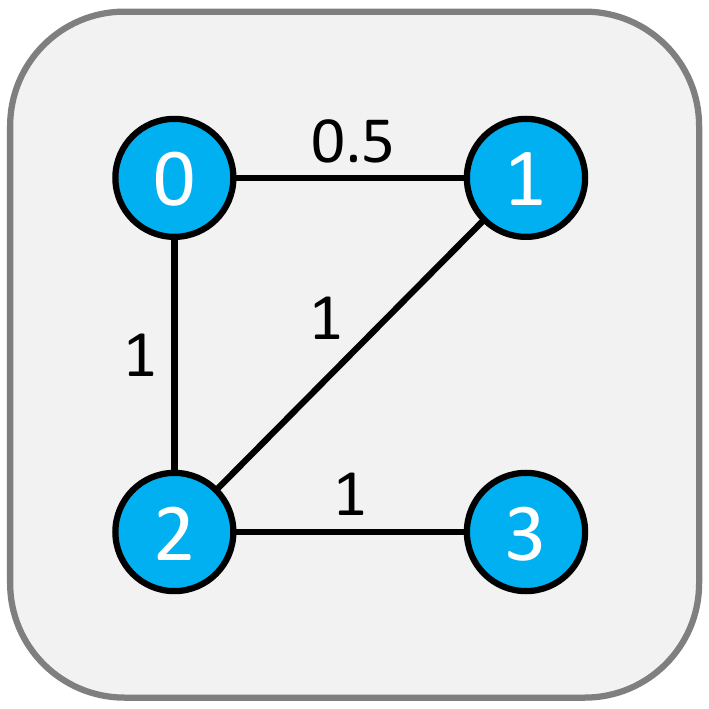}
    \hspace{2cm}
    \includegraphics[width=0.25\textwidth]{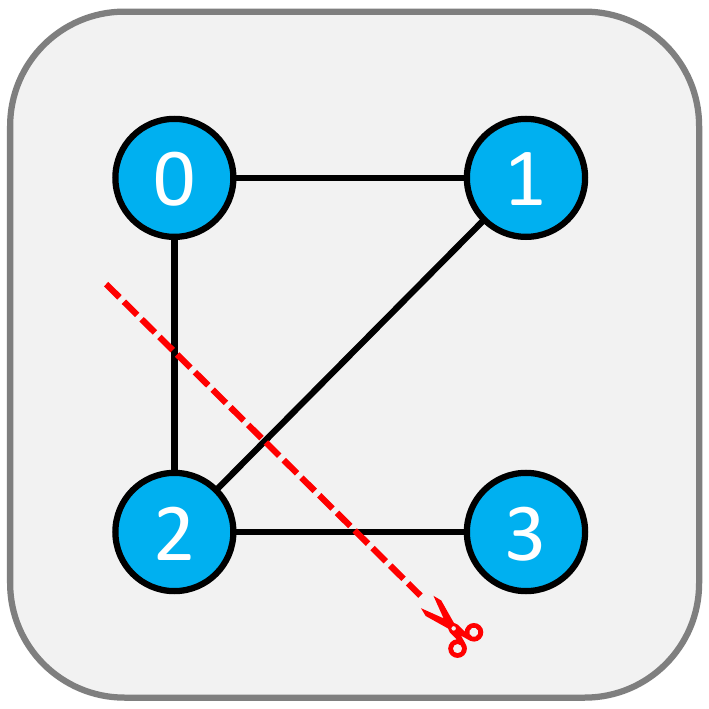}
    \caption{Example of the weighted max-cut problem. Given a weighted graph (left), the task is to find a partitioning which maximizes the sum of all weights on the cut (right). In this example, the best cut goes through all edges with weight one, partitioning the graph into the nodes $\{0,1,3\}$ and $\{2\}$.}
    \label{fig:max-cut-example}
\end{figure}

As a concrete example and a proof-of-principle of our method, we consider a weighted max-cut problem with four vertices.
The task of the weighted max-cut problem is to find a partitioning of the vertices in two complementary sets, such that the sum of all weights on the cut is maximized.
In Fig.~\ref{fig:max-cut-example} we show the graph and the optimal partitioning, which is given by dividing the vertices into the two sets $\{0,1,3\}$ and $\{2\}$ in our example.
The weighted max-cut problem can be formulated as a minimization problem of an Ising-Hamiltonian.
In our case the Hamiltonian is given by
\begin{equation}
    H_c = Z_2 Z_3+Z_0 Z_2+0.5\; Z_0 Z_1 +Z_1Z_2.\label{eq:qaoa_example}
\end{equation}
The optimal solution is given by the two bit-string (here and in the following we use the little-endian convention) $0100$ and $1011$, corresponding to the two sets mentioned above.

To solve the problem with QAOA we use the $p=1$ ansatz, which has two parameters:
\begin{equation}
    \ket{\psi} = e^{-i\frac{\beta}{2} (X_0+X_1+X_2+X_3) }e^{-i\frac{\gamma}{2} H_c } \ket{+}^{\otimes N}.
\end{equation}
From classical simulation we find the optimal angles $\gamma= -2.290$ and $\beta = -2.186$.

\begin{figure}
    \centering
    \includegraphics[trim=0 -6cm 0 0 ,width=0.35\textwidth]{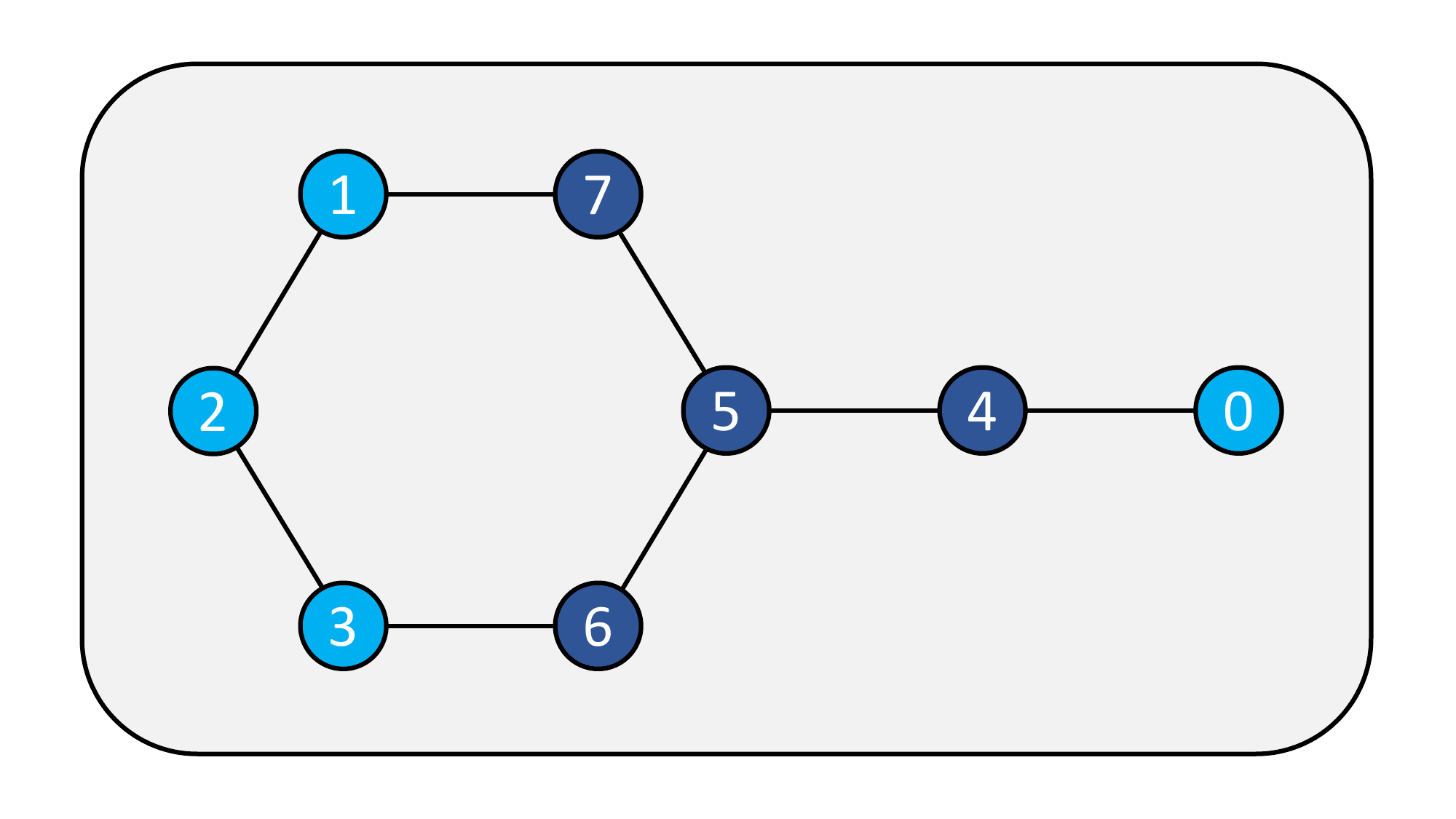}\quad
    \includegraphics[width=0.55\textwidth]{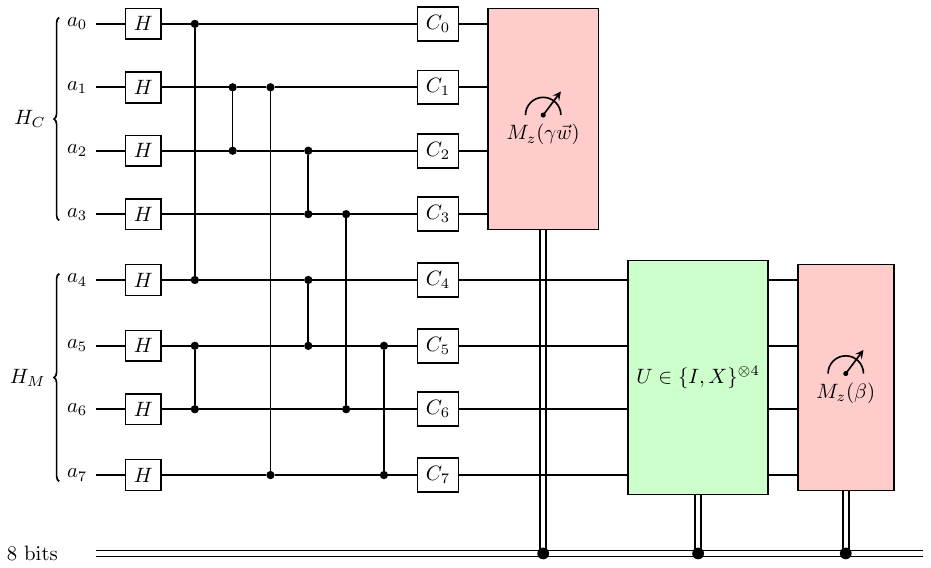}
    \caption{Left panel: graph state that implements the QAOA ansatz. The qubits $q_0-q_3$ correspond to the unitary implementing $\exp(-i\gamma/2 H_c)$, while qubits $q_4-q_6$ implement $\exp(-i\beta/2 H_m)$. Right panel: circuit which implements the graph state, together with the local unitaries which depend on the measurement outcome from the simulation of the main qubits, and the adaptive measurement instructions.}
    \label{fig:QAOAGraphCircuit}
\end{figure}

We first derive the pattern which implements this unitary, by mapping the two operators $e^{-i\frac{\gamma}{2} H_c }$ and $e^{-i\frac{\beta}{2} (X_0+X_1+X_2+X_3) }$ to stabilizer states.
Since both operations do not commute, this will introduce adaptive measurements.
We then simulate the main qubits and identify the graph shown in the left panel of Fig.~\ref{fig:QAOAGraphCircuit} as the full pattern, that implements the variational QAOA ansatz.

\begin{figure}
    \centering
    \includegraphics[width=0.95\textwidth]{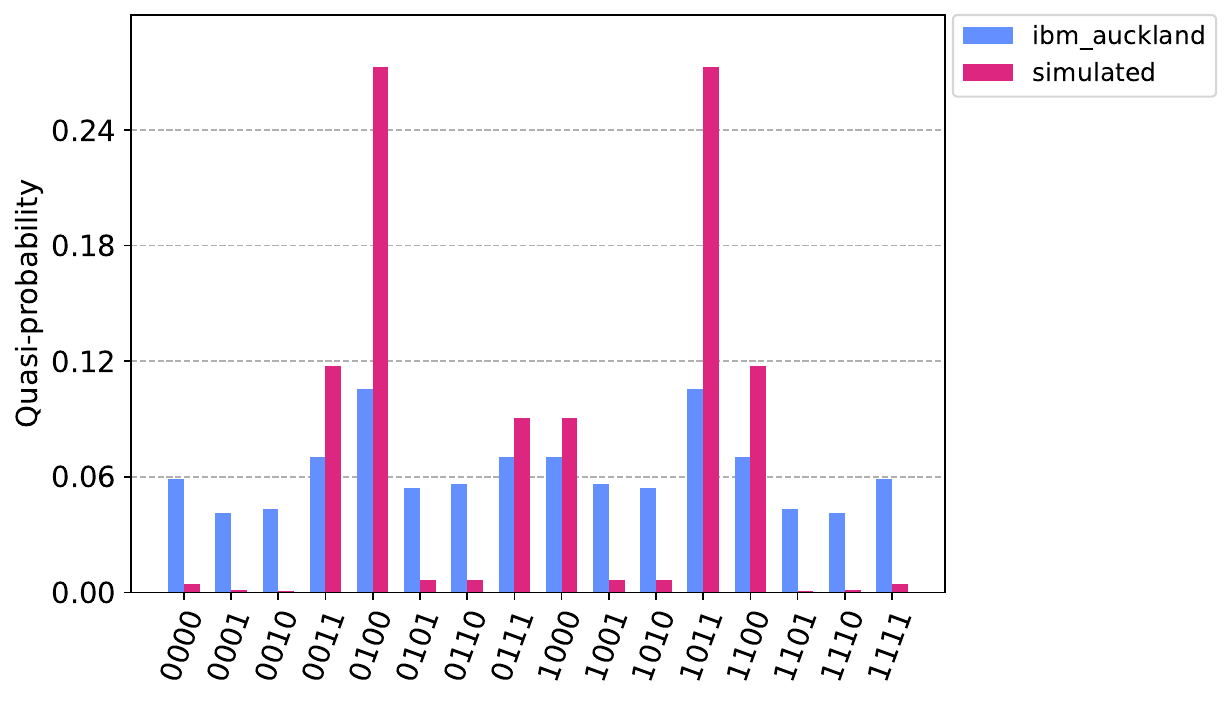}
    \caption{QAOA counts on \texttt{ibm\_auckland} using dynamic circuits and ideal, simulated counts. As expected, in the ideal distribution the correct solutions $0100$ and $1011$ are measured most often. The normalized fidelity (cf.~Eq.~\eqref{eq:fidelity}) between the ideal and simulated distribution is $F=0.28$.}
    \label{fig:qaoa_dynamic_circ}
\end{figure}

We ran the circuit corresponding to this pattern (right panel of Fig.~\ref{fig:QAOAGraphCircuit}) on the $27$-qubit quantum computer \texttt{ibm\_auckland}, using dynamic circuits.
The QAOA distribution, which is obtained by correcting the pre-simulated counts of the main qubits depending on the counts of the ancilla circuits, is shown in Fig.~\ref{fig:qaoa_dynamic_circ}.
We compare the results from \texttt{ibm\_auckland} with ideal results from simulation.
As expected, we find with highest probability the correct bit-string $0100$ and $1011$ in the simulated as well as the measured distributions.
However, due to hardware errors, the measured distribution differs from the ideal one.

To quantify the error, we calculate the normalized fidelity between the two distributions~\cite{lubinski2023applicationoriented}
\begin{equation}
    F\left(P_{\rm ideal}, P_{\rm measured}\right) = \frac{ F_H\left(P_{\rm ideal},P_{\rm measured}\right)-F_H\left(P_{\rm ideal},P_{\rm depol}\right)}{1-F_H\left(P_{\rm ideal},P_{\rm depol}\right)},\label{eq:fidelity}
\end{equation}
where $F_H$ denotes the Hellinger fidelity and $P_{\rm depol}$ corresponds to an uniform distribution, which would be measured on a completely depolarized device.
In our experiment, we find a fidelity of only $F=0.28$.

We believe that the main source of error is due to the use of dynamic circuits and measurement errors, which affect the whole outcome, if they occur during a mid-circuit measurement.
Note, that using dynamic circuits is a fairly new feature on IBM-hardware.
For the next application, the VQE, we therefore construct circuits, in which we avoid mid-circuit measurements by designing an ansatz with commuting operators only.

\subsection{VQE and the Unitary Coupled Cluster ansatz\label{sec:ucc}}
Quantum chemistry is often discussed as one of the most promising fields in which quantum computing could have a big impact.
The aim of ab-initio quantum chemistry is the calculation of molecular properties, such as their energies or polarization.
In second quantization, the molecular Hamiltonian is typically expressed in terms of fermionic annihilation and excitation operators,
\begin{equation}
    H = \sum_{p,q} h_{pq} a^{\dagger}_p a_q + \sum_{p,q,r,s} h_{pqrs}a^{\dagger}_p a^{\dagger}_q a_r a_s,\label{eq:molecular_hamiltonian}
\end{equation}
where $a_p$ ($a_p^{(\dagger)}$) annihilates (creates) an electron in the spin-orbital $p$.
An important task in quantum chemistry is the determination of the molecular ground state energy, which is given by the minimum of Hamiltonian \eqref{eq:molecular_hamiltonian}.
To achieve this goal using Quantum Computers, the Variational Quantum Eigensolver (VQE) has been thoroughly studied in the past decade~\cite{peruzzo2014variational,mcclean2016theory,TILLY20221,ryabinkin2018qubit,Xia_2021,nam2020ground}.

The unitary-coupled cluster (UCC) ansatz is among the most popular VQE-ansätze and is defined as
\begin{equation}
    \ket{\psi} = e^{i \sum_n T_n} \ket{\Phi}_0,
\end{equation}
where $\ket{\Phi}_0$ is the reference state (usually the Hartree-Fock ground state) and $T_n$ denote the $n$-th cluster operator -- usually these are truncated at second order, where $T_1$ and $T_2$ are given by
\begin{align}
    T_1 &= \sum_{\substack{i\in {\rm virt.}\\a\in {\rm occ.}}} \theta_a^i (a_i^{\dagger} a_a-a_a a_i^{\dagger}),\nonumber \\
    T_2 &= \sum_{\substack{i,j\in {\rm virt.}\\a,b\in {\rm occ.}}} \theta_{ab}^{ij} (a_i^{\dagger}a_j^{\dagger}a_a a_b-a_b^{\dagger}a_a^{\dagger}a_j a_i),
\end{align}
where virt.~(occ.)~denotes the set of virtual (occupied) orbitals.
Using the Jordan-Wigner mapping, those operators can be mapped to qubit operators via
\begin{align}
    a_n^{\dagger} \rightarrow Z_0 Z_1 \cdots Z_{n-1} \frac{X_n + i Y_n}{2}.
\end{align}
For the double-excitations $T_2$, this substitution leads to a sum of eight Pauli-strings.
In the literature these operators are often simplified by neglecting all $Z_i$ terms.
The operators in this approximation are called qubit excitation operators~\cite{ryabinkin2018qubit,Xia_2021}.
For instance, the qubit excitation corresponding to a fermionic double excitation operator in the Jordan-Wigner mapping is given by:
\begin{align}
    U_{ijab}=&e^{i\frac{\theta}{8}X_i Y_j X_a X_b}e^{i\frac{\theta}{8} Y_i X_j X_a X_b}e^{i\frac{\theta}{8} Y_i Y_j Y_a X_b}e^{i\frac{\theta}{8} Y_i Y_j X_a Y_b}\nonumber\\
    \times& e^{-i\frac{\theta}{8} X_i X_j Y_a X_b}e^{-i\frac{\theta}{8} X_i X_j X_a Y_b}e^{-i\frac{\theta}{8} Y_i X_j Y_a Y_b}e^{-i\frac{\theta}{8} X_i Y_j Y_a Y_b}.\label{eq:qubit_ucc}
\end{align}

\subsubsection{Measurement-pattern for double excitations}

\begin{figure}
    \centering
    \includegraphics[width=\textwidth]{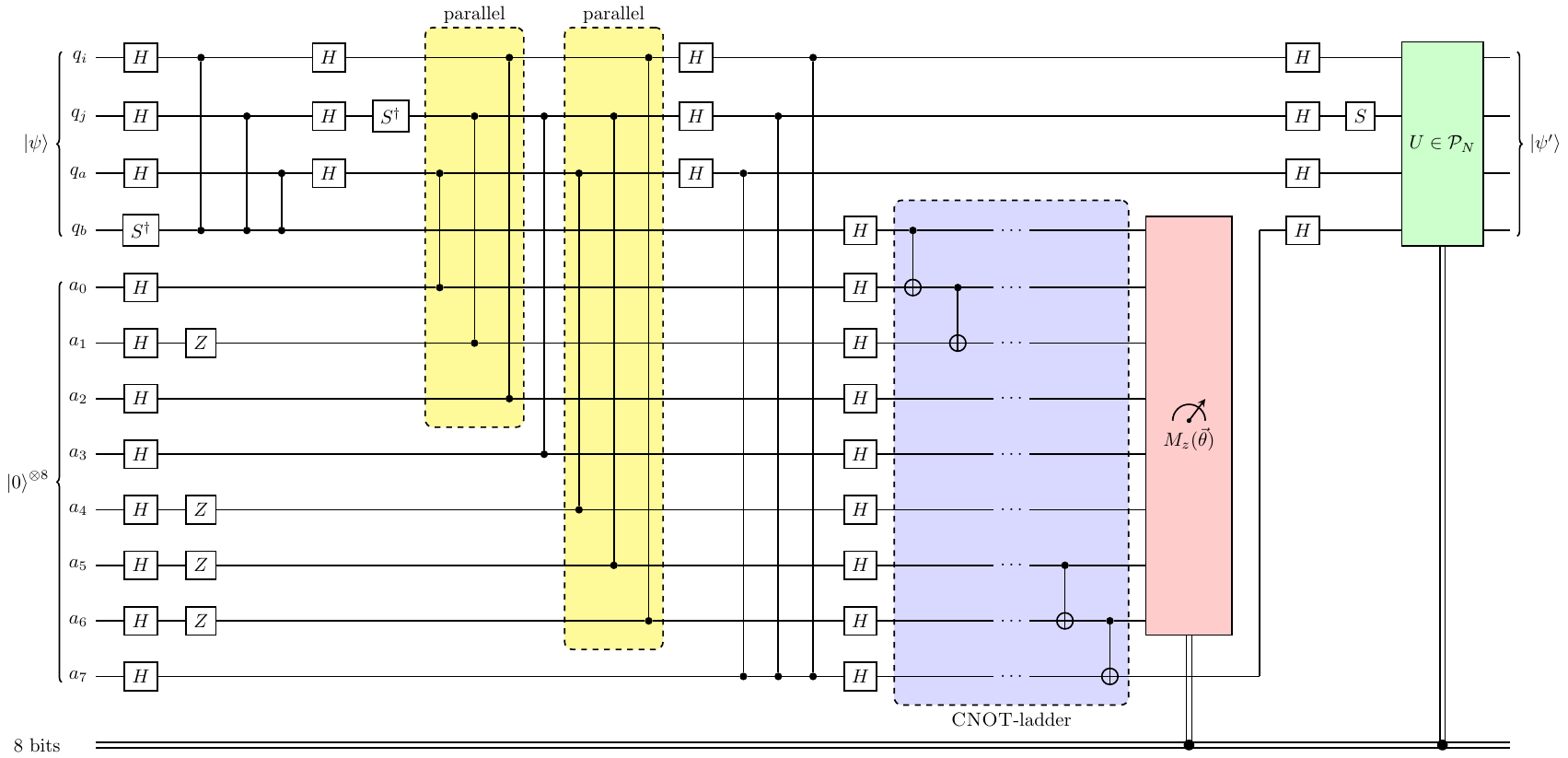}
    \caption{Measurement-pattern for the parallel application of the qubit double-excitation operator. The rotation angle for the measurement is given by $-\theta$ for qubits $q_b$, $a_0$, $a_1$ and $a_3$ and $\theta$ for qubits $a_2$, $a_4$, $a_5$ and $a_6$.}
    \label{fig:BBUCC}
\end{figure}

Next, we show how the qubit excitation in Eq.~\eqref{eq:qubit_ucc} (after rescaling $\theta \to 4 \theta$ for convenience) could be implemented in our protocol as a measurement pattern.
This operator has also been considered in Ref.~\cite{nam2020ground}, in which its circuit has been derived and optimized using the \textit{star} layout. This choice of layout is suited for our algorithm as the $R_z$-gates for the eight different Pauli strings in Eq.~\eqref{eq:qubit_ucc} can all be carried out on the same qubit (thus no initial ancilla is required).
In its optimized form, the circuit has $13$ CNOT-gates in total and is shown in the upper panel of Fig.~\ref{fig:BBUCC}.
We use it as our starting point to derive a measurement pattern.
The circuit obtained through our conversion algorithm is depicted in the lower panel of Fig.~\ref{fig:BBUCC}. 
According to Eq.~\eqref{eq:final_correction}, the final Pauli correction is given by 
\begin{multline}
    U = (X_i X_j X_a Y_b)^{s_1} \cdot (X_i X_j Y_a X_b)^{s_2} \cdot (X_i Y_j Y_a Y_b)^{s_3} \cdot (Y_i Y_j Y_a X_b)^{s_4} \\  \cdot (Y_i X_j Y_a Y_b)^{s_5} \cdot (Y_i X_j X_a X_b)^{s_6} \cdot (Y_i Y_j X_a Y_b)^{s_7} \cdot (X_i Y_j X_a X_b)^{s_8}.
\end{multline}

The circuit depth in terms of entangling gates is $17$. 
However, by only parallelizing the CNOT-ladder, the depth can be reduced to 11, which is already shorter than the conventional gate-based approach. 
As already discussed in Sec.~\ref{sec:ParallelClifford}, the depth of the Clifford circuit could be further reduced to a constant depth. 

\subsubsection{Proof of principle: Ground-state energy estimation of the H\textsubscript{2}O molecule}\label{sec:H2O_experiment}

\begin{figure}
    \centering
    \includegraphics[width=\textwidth]{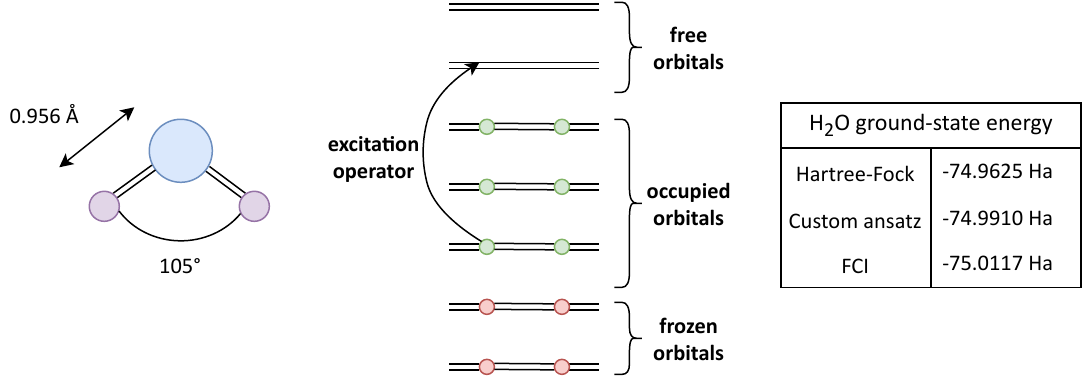}
    \caption{Left panel: Molecular geometry of the H\textsubscript{2}O molecule. Middle panel: Molecular orbitals contributing our ansatz using the \texttt{sto-3g} basis. We freeze the four electrons occupying the orbitals with the lowest energy and consider only excitations for the remaining six. Right panel: Ground-state energy of the H\textsubscript{2}O in the Hartree-Fock approximation, using our variational ansatz with frozen electrons (with optimized parameters) and the FCI energy, which is obtained by estimating the lowest eigenvalue of the full Hamiltonian.}
    \label{fig:h2o_orbital}
\end{figure}

In this section we use our techniques to lower the quantum-resource requirements of the VQE to estimate the ground-state energy of the H\textsubscript{2}O molecule on a quantum computer.
Our starting point is the electronic-structure Hamiltonian in the minimal \texttt{sto-3g} basis, which we derive using \texttt{openfermion}~\cite{McClean_2020} together with \texttt{pyscf}~\cite{https://doi.org/10.1002/wcms.1340}.
The full Hamiltonian consists of $14$ spin-orbitals, which are occupied by ten electrons.
In order to simplify the problem at hand, we freeze the four spin-orbitals with the lowest energy, such that we only deal with six electrons distributed over ten orbitals in our ansatz. A schematic overview of this approach is provided in Fig.~\ref{fig:h2o_orbital}.

Next, we choose our variational ansatz for the VQE.
The full UCC-ansatz in this case would result into too complicated patterns for current quantum hardware.
We therefore use a simplified ansatz, defined by an operator pool of Pauli strings, which is inspired by the qubit excitations.
This ansatz is known as qubit-ADAPT-VQE in the literature~\cite{PRXQuantum.2.020310}.
As shown in Sec.~\ref{sec:Parallelism}, we can parallelize all operations, if they commute.
We therefore aim to build our operator pool from commuting operators only.
The full pool, consisting of nine operators built from Pauli strings, is shown in Table~\ref{tab:pauli_exc}.

\begin{figure}
    \centering
    \includegraphics[width=\textwidth]{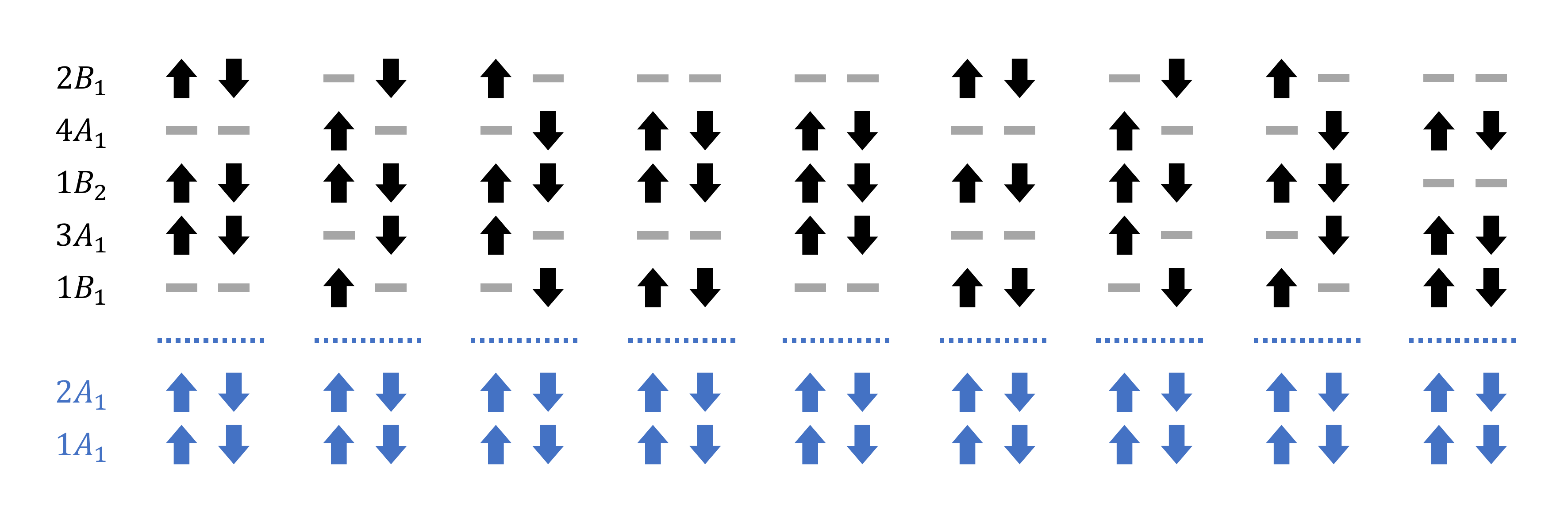}
    \caption{The nine most important excitations contributing to the ground-state energy of the H\textsubscript{2}O molecule following Ref.~\cite{nam2020ground}, in the approximation of the four inner-most electrons being frozen. Figure adapted from~\cite{nam2020ground}.}
    \label{fig:excitations}
\end{figure}

Our ansatz is inspired by the ansatz from Ref.~\cite{nam2020ground}, in which the most important qubit excitations were chosen -- instead of using the full qubit excitation operators consisting of eight Pauli strings each, we only choose one string per excitation.
Following Ref.~\cite{nam2020ground}, the most important excitations using the frozen core approximation, are shown in Fig.~\ref{fig:excitations}.
Our ansatz is then given by:
\begin{equation}
    \ket{\psi} = e^{-\frac{i}{2} \sum_n \theta_n P_n} \ket{\Phi}_\text{HF},
\end{equation}
where $\ket{\Phi}_\text{HF} \equiv \ket{0000111111}$ is the Hartree-Fock ground state and $\theta_n$ denotes the  $n$-th variational parameter.

\begin{figure}
    \centering
    \includegraphics[width=\textwidth]{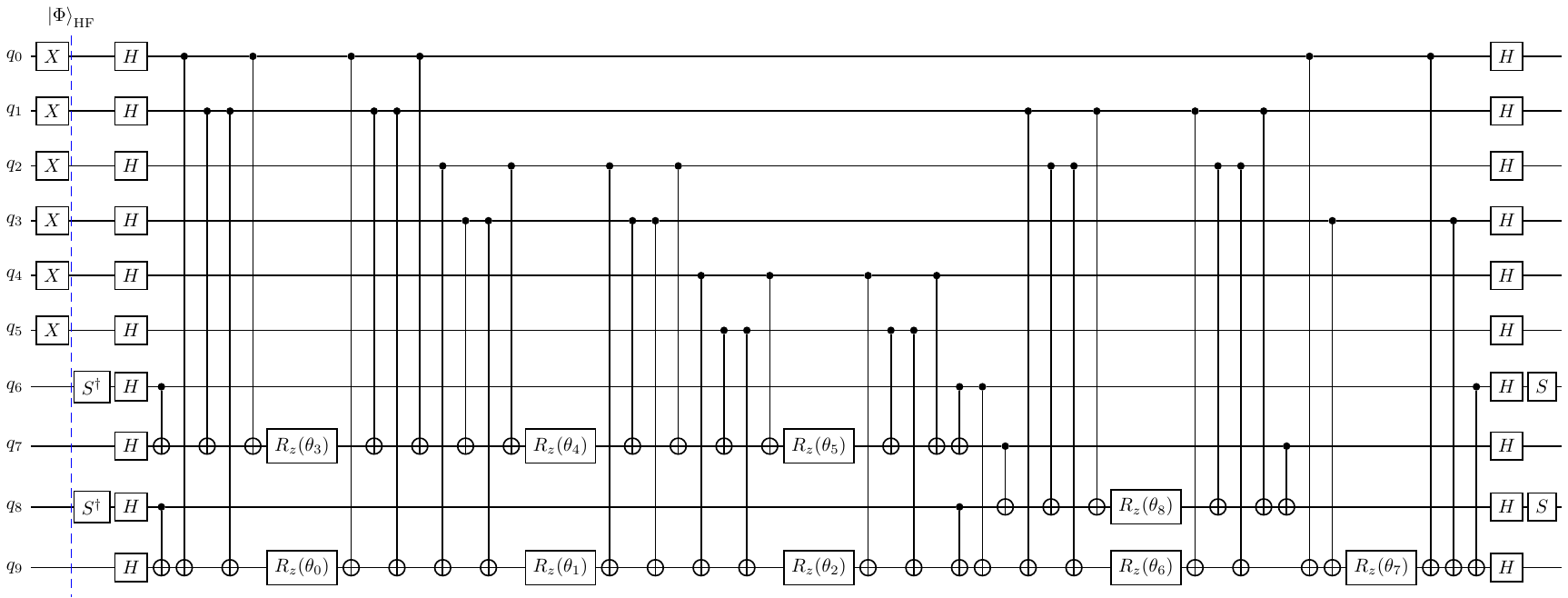}
    \caption{VQE-ansatz in the circuit model. To measure the expectation of one of the $14$ Hamiltonians from Eq.~\eqref{eq:ham_decomp} a specific Clifford circuit must be appended at the end~\cite{Crawford2021efficientquantum,gokhale2019minimizing}. This leads to high-depth circuits, which we reduce to shallow-depth pattern by mapping the ansatz to graph states.}
    \label{fig:VQE_ciruit}
\end{figure}

We first optimize our ansatz by performing a classical simulation of the circuit.
The optimal parameters leading to a minimal energy of $-74.9910$ Ha are shown in Tab.~\ref{tab:pauli_exc}.
In Fig.~\ref{fig:VQE_ciruit} we show the quantum circuit corresponding to this ansatz.

The Hamiltonian in our approximation consists of $251$ terms in total.
As observed in Refs.~\cite{Crawford2021efficientquantum,gokhale2019minimizing}, the sampling overhead of measuring the expectation value of such Hamiltonian can be reduced significantly by measuring commuting operators simultaneously.
Using the \texttt{qiskit} command \texttt{group\_commuting()} to identify commuting groups of a given operator, we find a partitioning of the Hamiltonian into $14$ groups of operators, which can be measured simultaneously.
More details to this decomposition can be found in Appendix~\ref{sec:ham_decomp}.
Our full Hamiltonian reads:
\begin{equation}
    H = -72.2129 {\rm\; Ha} + \sum_{n=1}^{14} H_n.\label{eq:ham_decomp}
\end{equation}
In order to measure the expectation value of each Hamiltonian $H_n$, we use the techniques outlined in Ref.~\cite{gokhale2019minimizing} to derive the Clifford circuits, which diagonalize a given set of commuting operators.

\begin{table}[t!]
    \centering
    \begin{tabular}{c|c}
    Operator $P_n$ & Parameter $\theta_n$\\\hline
      $X_0X_1Y_8X_9$   & $-0.157$\\
      $X_2X_3Y_8X_9$   & $-0.080$\\
      $X_4X_5Y_8X_9$   & $-0.023$\\
      $X_0X_1Y_6X_7$   & $-0.078$\\
      $X_2X_3Y_6X_7$   & $-0.081$\\
      $X_4X_5Y_6X_7$   & $-0.054$\\
      $X_1X_2Y_6X_9$   & $\phantom{-}0.099$\\
      $X_0X_3Y_6X_9$   & $-0.067$ \\
      $X_1X_2X_7Y_8$   & $-0.065$
    \end{tabular}
    \caption{Operator-pool for the VQE to estimate the ground-state of the H\textsubscript{2}O molecule. We construct nine commuting Pauli operators to avoid mid-circuit measurements.}
    \label{tab:pauli_exc}
\end{table}

\begin{figure}
    \centering
    \includegraphics[width=0.5\textwidth]{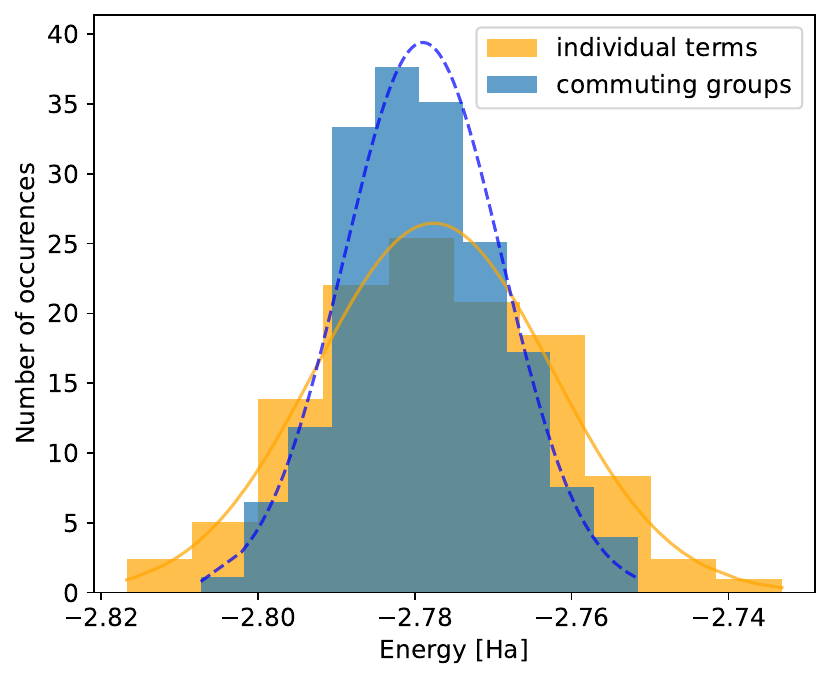}
    \caption{Comparison between the two sampling strategies. The data was obtained by simulating the quantum circuit shown in Fig.~\ref{fig:VQE_ciruit} (with optimized parameters) in \texttt{qiskit} $500$ times. For the case of measuring each term of the Hamiltonian individually, we use a total amount of roughly $250k$ shots in the simulation, while for the other approach we used only $70k$. There is an advantage by measuring commuting groups simultaneously compared to measuring each term individually, since the standard deviation due to shot noise is smaller in this approach.}
    \label{fig:sampling_comparision}
\end{figure}

To showcase the reduction of sampling overhead in this approach, we simulate the quantum circuit shown in Fig.~\ref{fig:VQE_ciruit} using the \texttt{qiskit} simulator.
In Fig.~\ref{fig:sampling_comparision} we compare the two strategies of estimating the expectation value of the Hamiltonian measuring each term individually vs.~using a partitioning into commuting groups.
In the first approach we measure each term of the Hamiltonian with $1000$ shots individually and in the second we use the technique of measuring each of the $14$ groups with $5000$ shots per group simultaneously.
We repeated the simulation $500$ times to collect sufficiently statistics to perform a fit with a normal distribution to estimate the variance of the expectation value due to shot noise for both strategies.
While in the first approach we obtain a standard deviation of $\sigma \approx 0.015$ using approximately $250k$ shots in total, the second approach leads to results with less shot noise ($\sigma \approx 0.01$) using only $70k$ shots.
This shows that using commuting groups of operators indeed gives a benefit for the measurement of the Hamiltonian in this example.

In the following, we construct the variational quantum circuit for each of the $14$ groups by concatenating the ansatz circuit (Fig.~\ref{fig:VQE_ciruit}) with the corresponding Clifford circuit, that diagonalizes all operators in a given group.
We then use our algorithm to derive measurement patterns, which implement the whole quantum operation.
After classically simulating the main qubits, we end up with very simple graph states after using the optimization procedure outlined in Sec.~\ref{sec:optimizing_graph_states}.

\begin{figure}
    \centering
    \includegraphics[width=0.95\textwidth]{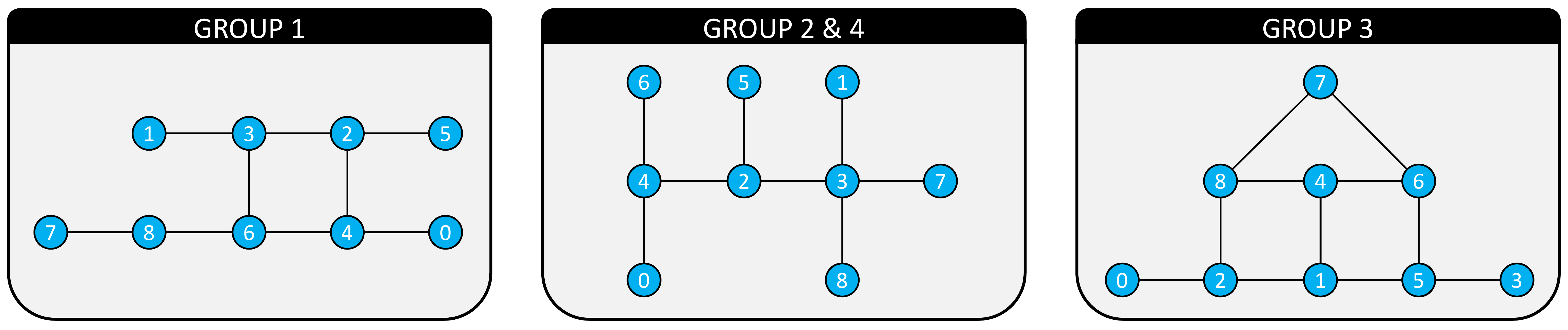}
    \caption{Graph states corresponding to the measurement of the first four terms of the Hamiltonian~\eqref{eq:ham_decomp}. Each graph represents $32$ locally-equivalent states contributing to the expectation value.}
    \label{fig:circuit_to_graph_states_h2o}
\end{figure}

Next, we benchmark our ansatz using the optimized parameters and ran the corresponding circuits on the $27$-qubit quantum computer \texttt{ibm\_hanoi}.
As a proof-of-principle we only ran those circuits associated with the energies of the first four terms of the Hamiltonian.
We show the graph states corresponding to these four terms in Fig.~\ref{fig:circuit_to_graph_states_h2o}.
We report our final results in Tab.~\ref{tab:measured_energies}.
For each group, we find that $32$ locally-equivalent stabilizer states have to be prepared.
Accordingly, we ran $32$ circuits per group using $4k$ shots, such that each expectation value was measured with a total budget of $128k$ shots.
The error for the raw data shown in Tab.~\ref{tab:measured_energies} (third column) are estimated by repeating each experiment eight times and calculating the standard deviation.

As can be seen from Tab.~\ref{tab:measured_energies}, the energies calculated with the raw data-points are far-off the ideal values, such that the use of error-mitigation is imperative.
We mitigate read-out error using the \texttt{M3} package~\cite{PRXQuantum.2.040326}, for which we ran all calibration circuits with $100\;000$ shots.
Additionally to the read-out error correction we used randomized-compiling~\cite{PhysRevA.94.052325}, dynamical decoupling~\cite{PhysRevA.58.2733} and zero-noise-extrapolation (ZNE)~\cite{PhysRevLett.119.180509,giurgica2020digital} to increase the precision.
The mitigated values are shown in the last column of Tab.~\ref{tab:measured_energies}.
The reported error follows from the ZNE fit parameter.

It is important to note that the applicability of ZNE is not evident here, as observables are measured classically on the main qubits and only later corrected through the ancilla outcomes. 
In Appendix \ref{sec:error_mit} we prove that the expectation value of an arbitrary Pauli string $\mathcal{P}_q$, acting purely on the main qubits, can be expressed as a superposition of expectation values of an auxiliary Pauli-$Z$ string $\mathcal{Z}_a$ acting on the ancilla register, that is
\begin{equation}
     \braket{\mathcal{P}_q} =\frac{1}{N} \sum_{n} (-1)^{s_n}
    \prescript{}{n}{\bra{\psi_a}}\mathcal{Z}_a \ket{\psi_a}_n,
\end{equation}
where $N$ is the number of distinct classical measurement outcomes, $\ket{\psi_a}_n$ are the ancilla states corresponding to the classical measurement outcome $n$ and $s_n\in\{0,1\}$ ensures the correct phase. From this form it is clear that by amplifying the noise on the ancilla systems $\ket{\psi_a}_n$, ZNE can be performed on the main system. More details on the error-mitigation and the data acquisition can be found in Appendices~\ref{sec:error_mit} and \ref{sec:data_vqe}.

Note, that comparing the ideal expectation value with the mitigated ones, we achieve an absolute accuracy of roughly $0.01$ Ha. 
While we accomplished promising improvements over the unmitigated results, our mitigated results are still not within the range of ``chemical accuracy'' of $0.001$ Ha~\cite{RevModPhys.92.015003}. 
To be able to make the full calculation with all $14$ groups to such accuracy, more sophisticated error-mitigation~\cite{van2023probabilistic,gupta2023probabilistic} will be necessary together with a longer access to the quantum computer to get better statistics.
However, as a proof-of-principle, we believe that this experiment shows that our techniques are very promising for future research.

\begin{table}[t!]
    \centering
    \begin{tabular}{c|c|c|c}
        Hamiltonian & Exp.~val.~ideal & Exp.~val.~measured & Exp.~val.~mitigated\\ \hline
        $H_1$ &  $-0.2565$ & $-0.4396\pm 0.0014$ &  $-0.2698\pm 0.0024$ \\
        $H_2$ & $-0.2546$ & $-0.1852\pm 0.0008$ & $-0.2460\pm 0.0008$ \\
        $H_3$ & $\phantom{-}1.0223$ & $\phantom{-}0.7691\pm 0.0026$ & $\phantom{-}1.0143\pm 0.0028$ \\
        $H_4$ & $-0.1223$ & $-0.0924\pm 0.0007$ & $-0.1223 \pm 0.0008$ \\
    \end{tabular}
    \caption{Expectation values of the Hamiltonians (and their respective standard deviations) measured on the quantum computer in comparison with the ideal ones. As a proof-of-principle  we only measured the first four Hamiltonians.}
    \label{tab:measured_energies}
\end{table}

\section{Conclusion and outlook}
In this work, we introduced an algorithm that allows the mapping of a sequence of unitaries generated by Pauli strings in the quantum circuit model to a measurement pattern by introducing one ancilla qubit per unitary.
We showed that in the case of commuting operators these patterns can be parallelized leading to a constant-depth quantum operation.
This result is useful in the NISQ-era, in which quantum circuits have to be as shallow as possible, as well as when fault-tolerant quantum computers are available and one needs to be able to perform unitary operations fast in parallel.
The pattern always consists of three layers: a Clifford part, the measurement of the ancilla qubits, and a correction layer that consists of local Pauli operations.
Furthermore, we showed that by compressing the Clifford part to an LC-equivalent graph state and simulating the main qubits classically, we can significantly reduce the complexity of a given quantum circuit. We further optimized our graph states using simulated annealing to minimize the hardware requirements for state preparation.

We first applied the algorithm to the QAOA. 
Despite obtaining correct results in the simulation, our experiments on quantum hardware achieved a rather poor fidelity of $F=0.28$, mostly due to errors caused by mid-circuit measurements. 
To overcome this hurdle on current NISQ-hardware, one could try to design ansätze, that do not require mid-circuit measurements, as we have done for the VQE.
One possible direction could be to restrict the mixer-operator space to Clifford operators, i.e.,~$\{I, H, S\}$, thus removing the necessity for mid-circuit measurements. 
As has been shown in \cite{munoz2023low}, restricting the QAOA ansatz to the pure Clifford manifold already provides good approximate solutions to the max-cut problem.
Combining this ansatz with a measurement-pattern consisting of commuting operators only, might be a possible road to find more efficient ansätze.

As a second example, we applied our mapping technique to the VQE to compute the ground-state energy of H\textsubscript{2}O. 
Here, we specifically designed an ansatz of fully-commuting operators to avoid mid-circuit measurements. 
This approach significantly outperforms the standard circuit model approach in terms of circuit depth.
We showed that running our circuits on current IBM-hardware, we were able to extract expectation values with an absolute accuracy of roughly $0.01$ Ha. 
This was only possible using error-mitigation techniques to boost the precision of our results.
We showed, how such error-mitigation techniques can be incorporated in our formalism by performing zero-noise extrapolation using local gate folding.

We believe that our mapping techniques will provide a useful alternative compared to the standard circuit model for future NISQ-algorithms.
It is straightforward to apply our techniques to other cases, such as quantum circuits following from trotterized time evolution or other VQE ansätze in different scenarios.
There are several directions for future research.
First, we want to study how graph states can be implemented more efficiently on a quantum computer by optimizing the graph with respect to its topology.
By using local complementation and Pauli measurements, hardware-efficient graphs could be implemented which are equivalent to the target state one wants to prepare.
Secondly, since the number of qubits on a quantum computer is limited, an important line of research is how ancilla qubits can be reused efficiently after their measurement. 
An interesting alternative approach could be to use circuit-cutting techniques \cite{tang2021cutqc} to cut the graph state into several partitions.
Here, it is important to cut the circuit corresponding to the measurement pattern at a location, such that the additional sampling overhead is minimal.
Last but not least, we want to study how probabilistic error cancellation \cite{van2023probabilistic, gupta2023probabilistic} can be incorporated in our framework.
\clearpage

\section{Acknowledgments}
We thank Kathrin König for sharing her code for zero-noise-extrapolation, which we used to perform the local gate-folding. 
We would also like to thank Jonas Jäger, Daniel Barragan-Yani, Paul Haubenwallner, Timon Scheiber and Johannes S. Mueller-Roemer for helpful comments on the manuscript.
This work was supported in part by the research project \textit{Zentrum für Angewandtes Quantencomputing} (ZAQC), which is funded by the Hessian
Ministry for Digital Strategy and Innovation and the Hessian Ministry of Higher Education, Research and the Arts (TNK and MH), and the research project \textit{Quantum Computing for Materials Science and Engineering} (QuantiCoM) of the DLR \textit{Quantencomputing Initiative} (QCI), which is funded by the Federal Ministry of Economic Affairs and Climate Action (BMWK) (TNK). We acknowledge the use of IBM Quantum services for this work. The views expressed are those of the authors, and do not reflect the official policy or position of IBM or the IBM Quantum team.

\clearpage
\printbibliography

\clearpage
\appendix

\section{Conditions for parallel measurements of the ancilla qubits} \label{app:ParallelismAdapativity}
In this Appendix we derive Eq.~\eqref{eq:correction_commutation}.
Consider two Pauli strings $\mathcal{P}$ and $\mathcal{\tilde P}$.
We want to derive under which conditions we can apply the measurement pattern implementing the unitary \begin{equation}
    U=e^{-\frac{i}{2}\tilde\theta\tilde{\mathcal{P}}}e^{-\frac{i}{2}\theta{\mathcal{P}}}
\end{equation}
with parallel, i.e., non-adaptive, measurements.
We know, that the implementation of the first pattern, which applies the unitary $e^{-\frac{i}{2}\theta{\mathcal{P}}}$, leads to a Pauli correction given by $\mathcal{P}^s$, where $s$ is the measurement outcome of the ancilla.
This correction needs to be shifted through the pattern implementing the second unitary $e^{-\frac{i}{2}\tilde\theta\tilde{\mathcal{P}}}$.
This can be achieved as follows.

First, we decompose the matrix exponential
\begin{equation}
     e^{-\frac{i}{2}\tilde\theta\tilde{\mathcal{P}}} = \left[\cos\left(\frac{\tilde\theta}{2}\right) -i \sin\left(\frac{\tilde\theta}{2}\right)\tilde{\mathcal{P}}\right].
\end{equation}
We then shift $\mathcal{P}^s$ through the second unitary by inserting an identity,
\begin{equation}
    e^{-\frac{i}{2}\tilde\theta\tilde{\mathcal{P}}} \mathcal{P}^s = \mathcal{P}^s \left[\cos\left(\frac{\tilde\theta}{2}\right) -i \sin\left(\frac{\tilde\theta}{2}\right)\mathcal{P}^s\tilde{\mathcal{P}}\mathcal{P}^s\right].
    \label{eq:RefStep}
\end{equation}
In case that $[\mathcal{P}, \tilde{\mathcal{P}}]=0$, we have $\mathcal{P}^s\tilde{\mathcal{P}}\mathcal{P}^s = \tilde{\mathcal{P}}$, and thus obtain the trivial commutation relation

\begin{equation}
    e^{-\frac{i}{2}\tilde\theta\tilde{\mathcal{P}}} \mathcal{P}^s = \mathcal{P}^s  e^{-\frac{i}{2}\tilde\theta\tilde{\mathcal{P}}}. 
\end{equation}
However, if $[\mathcal{P}, \tilde{\mathcal{P}}]\neq 0$, we have $\mathcal{P}^s\tilde{\mathcal{P}}\mathcal{P}^s = (-1)^s\tilde{\mathcal{P}}$. Then, we can rearrange Eq.~\eqref{eq:RefStep} as follows:
\begin{equation}
    \begin{split}
        e^{-\frac{i}{2}\tilde\theta\tilde{\mathcal{P}}} \mathcal{P}^s &= \mathcal{P}^s \left[\cos\left(\frac{\tilde\theta}{2}\right) -i \sin\left(\frac{\tilde\theta}{2}\right)(-1)^s\tilde{\mathcal{P}}\right] \\
        &= \mathcal{P}^s \left[\cos\left((-1)^s\frac{\tilde\theta}{2}\right) -i \sin\left((-1)^s\frac{\tilde\theta}{2}\right)\tilde{\mathcal{P}}\right] \\
        &= \mathcal{P}^s  e^{-\frac{i}{2}(-1)^s\tilde\theta\tilde{\mathcal{P}}}.
    \end{split}
\end{equation}
Consequently, by applying our protocol to non-commuting strings, adaptive rotations are introduced, leading to adaptive measurements.

\section{Simulated annealing on graph states} \label{app:SimulatedAnnealing}
In this Appendix we summarize the simulated annealing algorithm we used to simplify graph states.
The goal is to minimize a cost function $f:D\to \mathbb{R}$ on a solution space $D$. The solution space is the set of equivalent graph states in our case. 

\begin{enumerate}
    \item \textit{Initialization} \\
    Select an initial solution $g\in D$ and a monotonously falling sequence of (positive) temperatures $T_i$.
    
    \item \textit{Local change} \\
    Select a neighbor $\tilde g$ of $g$. On our set of graph states, neighbors of $g$ are defined as graphs, which are related to $g$ by a local complementation at a single node. We construct such a neighbor by locally complementing a random node of $g$. 
    
    \item \textit{Select} \\
    If $f(\tilde g)\leq f(g)$, set $g=\tilde g$. Else, set $g=\tilde g$ with a probability of 
    
    \begin{equation}
        P(g,\tilde g) = \exp\left(-\frac{f(\tilde g)-f(g)}{T_i}\right).
    \end{equation}
    
    \item \textit{Update} \\
    If $f(g)$ is better than the previous best solution, update it. 
    
    \item \textit{Increment} \\
    Set $i = i + 1$.
    
    \item \textit{Repeat} \\
    Repeat steps 2-5 until the final temperature is reached. 
\end{enumerate}

\section{Grouping the H\textsubscript{2}O Hamiltonian}
\label{sec:ham_decomp}
In this Appendix we show the first four groups of the Hamiltonian of H\textsubscript{2}O. As mentioned in Sec.~\ref{sec:H2O_experiment}, we used the built-in function \texttt{group\_commuting()} in \texttt{qiskit} to find the partitioning.
In Tab.~\ref{tab:group1_2}, we show the terms contributing to $H_1$ and $H_2$ and in Tab.~\ref{tab:group3_4} the terms contributing to $H_3$ and $H_4$.

Finding commuting groups of an Hamiltonian scales exponentially with the system size in general.
However, as explained in Ref.~\cite{gokhale2019minimizing}, for the electronic-structure problem one can perform a partitioning of the Hamiltonian which scales polynomially with the system size.

\begin{table}[H]
    \centering
    \begin{tabular}{|c|c|}\hline
    \multicolumn{2}{|c|}{$H_1$}\\\hline
    Pauli & Coeff.\\
    \hline
        $Y_0 X_1 X_2 Y_3 $ & $0.01176$\\$Y_0 Y_1 X_2 X_3 $ & $-0.01176$\\$X_0 X_1 Y_2 Y_3 $ & $-0.01176$\\$X_0 Y_1 Y_2 X_3 $ & $0.01176$\\$Y_0 Z_1 Y_2 Y_7 Z_8 Y_9 $ & $0.01186$\\$Y_0 Z_1 Y_2 X_7 Z_8 X_9 $ & $0.01186$\\$X_0 Z_1 X_2 Y_7 Z_8 Y_9 $ & $0.01186$\\$X_0 Z_1 X_2 X_7 Z_8 X_9 $ & $0.01186$\\$Y_1 Z_2 Y_3 Y_6 Z_7 Y_8 $ & $0.01186$\\$Y_1 Z_2 Y_3 X_6 Z_7 X_8 $ & $0.01186$\\$X_1 Z_2 X_3 Y_6 Z_7 Y_8 $ & $0.01186$\\$X_1 Z_2 X_3 X_6 Z_7 X_8 $ & $0.01186$\\$Y_6 X_7 X_8 Y_9 $ & $0.02894$\\$Y_6 Y_7 X_8 X_9 $ & $-0.02894$\\$X_6 X_7 Y_8 Y_9 $ & $-0.02894$\\$X_6 Y_7 Y_8 X_9 $ & $0.02894$\\$Y_0 Z_1 Y_2 X_6 Z_7 X_8 $ & $0.00076$\\$X_0 Z_1 X_2 Y_6 Z_7 Y_8 $ & $0.00076$\\$Y_1 Z_2 Y_3 X_7 Z_8 X_9 $ & $0.00076$\\$X_1 Z_2 X_3 Y_7 Z_8 Y_9 $ & $0.00076$\\$Z_0 Z_2 $ & $0.13797$\\$Y_0 Z_1 Y_2 Y_6 Z_7 Y_8 $ & $-0.00757$\\$X_0 Z_1 X_2 X_6 Z_7 X_8 $ & $-0.00757$\\$Z_1 Z_3 $ & $0.13797$\\$Y_1 Z_2 Y_3 Y_7 Z_8 Y_9 $ & $-0.00757$\\$X_1 Z_2 X_3 X_7 Z_8 X_9 $ & $-0.00757$\\$Z_6 Z_8 $ & $0.1126$\\$Z_7 Z_9 $ & $0.1126$\\$Z_4 $ & $0.48237$\\$Z_5 $ & $0.48237$\\$Z_4 Z_5 $ & $0.22004$\\
        \hline
    \end{tabular}
    \quad
    \begin{tabular}{|c|c|}\hline
    \multicolumn{2}{|c|}{$H_2$}\\\hline
    Pauli & Coeff.\\
    \hline
       $Y_0 X_1 X_3 Z_4 Z_5 Y_6 $ & $0.00715$\\$Y_0 Y_1 Y_3 Z_4 Z_5 Y_6 $ & $0.00715$\\$X_0 X_1 X_3 Z_4 Z_5 X_6 $ & $0.00715$\\$X_0 Y_1 Y_3 Z_4 Z_5 X_6 $ & $0.00715$\\$Y_0 Z_1 Z_2 Z_3 Z_4 Z_5 Y_6 Y_7 Z_8 Y_9 $ & $-0.01627$\\$Y_0 Z_1 Z_2 Z_3 Z_4 Z_5 Y_6 X_7 Z_8 X_9 $ & $-0.01627$\\$X_0 Z_1 Z_2 Z_3 Z_4 Z_5 X_6 Y_7 Z_8 Y_9 $ & $-0.01627$\\$X_0 Z_1 Z_2 Z_3 Z_4 Z_5 X_6 X_7 Z_8 X_9 $ & $-0.01627$\\$Y_1 X_2 X_3 Z_4 Z_5 Z_6 Z_7 Y_8 $ & $0.00064$\\$Y_1 Y_2 X_3 Z_4 Z_5 Z_6 Z_7 X_8 $ & $-0.00064$\\$X_1 X_2 Y_3 Z_4 Z_5 Z_6 Z_7 Y_8 $ & $-0.00064$\\$X_1 Y_2 Y_3 Z_4 Z_5 Z_6 Z_7 X_8 $ & $0.00064$\\$Y_2 Z_3 Z_4 Z_5 Z_6 X_7 X_8 Y_9 $ & $-0.01444$\\$Y_2 Z_3 Z_4 Z_5 Z_6 Y_7 X_8 X_9 $ & $0.01444$\\$X_2 Z_3 Z_4 Z_5 Z_6 X_7 Y_8 Y_9 $ & $0.01444$\\$X_2 Z_3 Z_4 Z_5 Z_6 Y_7 Y_8 X_9 $ & $-0.01444$\\$Y_0 Z_1 X_2 X_6 Z_7 Y_8 $ & $-0.00832$\\$X_0 Z_1 Y_2 Y_6 Z_7 X_8 $ & $-0.00832$\\$Z_0 Z_6 $ & $0.12496$\\$Z_2 Z_8 $ & $0.13512$\\
        \hline
    \end{tabular}

    \caption{Terms contributing to the Hamiltonians $H_1$ and $H_2$ (first two groups). All terms in one table commute with each other.}
    \label{tab:group1_2}
\end{table}

\begin{table}[H]
    \centering
\begin{tabular}{|c|c|}\hline
    \multicolumn{2}{|c|}{$H_3$}\\\hline
    Pauli & Coeff.\\
    \hline
       $Y_0 X_1 X_4 Y_5 $ & $0.0072$\\$Y_0 Y_1 X_4 X_5 $ & $-0.0072$\\$X_0 X_1 Y_4 Y_5 $ & $-0.0072$\\$X_0 Y_1 Y_4 X_5 $ & $0.0072$\\$Y_2 X_3 X_8 Y_9 $ & $0.01716$\\$Y_2 Y_3 X_8 X_9 $ & $-0.01716$\\$X_2 X_3 Y_8 Y_9 $ & $-0.01716$\\$X_2 Y_3 Y_8 X_9 $ & $0.01716$\\$Z_2 Z_9 $ & $0.15228$\\$Z_3 Z_8 $ & $0.15228$\\$Z_3 Z_9 $ & $0.13512$\\$Z_0 Z_1 $ & $0.1583$\\$Z_2 Z_3 $ & $0.19617$\\$Z_6 Z_7 $ & $0.14912$\\$Z_8 Z_9 $ & $0.15503$\\$Z_0 Z_4 $ & $0.15003$\\$Z_0 Z_5 $ & $0.15723$\\$Z_1 Z_4 $ & $0.15723$\\$Z_1 Z_5 $ & $0.15003$\\$Z_6 $ & $0.10364$\\$Z_7 $ & $0.10364$\\
        \hline
    \end{tabular}
    \quad
    \begin{tabular}{|c|c|}\hline
    \multicolumn{2}{|c|}{$H_4$}\\\hline
    Pauli & Coeff.\\
    \hline
       $Y_0 X_1 X_2 Z_3 Z_4 Z_5 Z_6 Y_7 $ & $-0.00715$\\$Y_0 Y_1 X_2 Z_3 Z_4 Z_5 Z_6 X_7 $ & $0.00715$\\$X_0 X_1 Y_2 Z_3 Z_4 Z_5 Z_6 Y_7 $ & $0.00715$\\$X_0 Y_1 Y_2 Z_3 Z_4 Z_5 Z_6 X_7 $ & $-0.00715$\\$Y_0 Z_1 Y_2 Y_3 Z_4 Z_5 Z_6 Z_7 Z_8 Y_9 $ & $0.00064$\\$Y_0 Z_1 Y_2 X_3 Z_4 Z_5 Z_6 Z_7 Z_8 X_9 $ & $0.00064$\\$X_0 Z_1 X_2 Y_3 Z_4 Z_5 Z_6 Z_7 Z_8 Y_9 $ & $0.00064$\\$X_0 Z_1 X_2 X_3 Z_4 Z_5 Z_6 Z_7 Z_8 X_9 $ & $0.00064$\\$Y_1 Z_2 Z_3 Z_4 Z_5 X_6 X_7 Y_8 $ & $-0.01627$\\$Y_1 Z_2 Z_3 Z_4 Z_5 Y_6 X_7 X_8 $ & $0.01627$\\$X_1 Z_2 Z_3 Z_4 Z_5 X_6 Y_7 Y_8 $ & $0.01627$\\$X_1 Z_2 Z_3 Z_4 Z_5 Y_6 Y_7 X_8 $ & $-0.01627$\\$Y_3 Z_4 Z_5 X_6 X_8 Y_9 $ & $0.01444$\\$Y_3 Z_4 Z_5 Y_6 Y_8 Y_9 $ & $0.01444$\\$X_3 Z_4 Z_5 X_6 X_8 X_9 $ & $0.01444$\\$X_3 Z_4 Z_5 Y_6 Y_8 X_9 $ & $0.01444$\\$Y_1 Z_2 X_3 X_7 Z_8 Y_9 $ & $-0.00832$\\$X_1 Z_2 Y_3 Y_7 Z_8 X_9 $ & $-0.00832$\\$Z_1 Z_7 $ & $0.12496$\\
        \hline
\end{tabular}
\caption{Terms contributing to the Hamiltonians $H_3$ and $H_4$.}
\label{tab:group3_4}
\end{table}

In order to measure the terms from these four groups simultaneously, we employ the algorithm of Ref.~\cite{gokhale2019minimizing}.
In the first step, we identify an operator basis for each group, from which all other operators can be constructed by multiplication.
For the four groups, we chose the following bases:
\begin{itemize}
    \item $H_1$: $Z_6 Z_8$,
        $Z_7 Z_9$,
        $Z_5$,
        $Z_4 Z_5$,
        $X_6 Y_7 Y_8 X_9$,
        $Y_0 Z_1 Y_2 Y_6 Z_7 Y_8$,
        $X_0 Z_1 X_2 X_6 Z_7 X_8$,
        $Y_1 Z_2 Y_3 Y_7 Z_8 Y_9$,
        $X_1 Z_2 X_3 X_7 Z_8 X_9$,

    \item $H_2$: $Y_1 Y_2 X_3 Z_4 Z_5 Z_6, Z_7 X_8$,
        $X_1 Y_2 Y_3 Z_4 Z_5 Z_6 Z_7 X_8$,
        $X_2 Z_3 Z_4 Z_5 Z_6 X_7 Y_8 Y_9$,
        $X_2 Z_3 Z_4 Z_5 Z_6 Y_7 Y_8 X_9$,
        $X_0 Z_1 Y_2 Y_6 Z_7 X_8$,
        $Z_0 Z_6$,
        $Z_2 Z_8$,
        
    \item $H_3$: $Z_3 Z_9 $,
        $Z_2 Z_3 $,
        $Z_8 Z_9 $,
        $Z_0 Z_5 $,
        $Z_1 Z_4 $,
        $Z_1 Z_5 $,
        $Z_6 $,
        $Z_7 $,
        $X_0 Y_1 Y_4 X_5 $,
        $X_2 Y_3 Y_8 X_9 $,
        
    \item $H_4$: 
        $X_3 Z_4 Z_5 X_6 X_8 X_9 $,
        $X_3 Z_4 Z_5 Y_6 Y_8 X_9 $,
        $Y_1 Z_2 X_3 X_7 Z_8 Y_9 $,
        $X_1 Z_2 Y_3 Y_7 Z_8 X_9 $,
        $Y_0 Z_1 Y_2 X_3 Z_4 Z_5 Z_6 Z_7 Z_8 X_9 $,
        $X_0 Z_1 X_2 X_3 Z_4 Z_5 Z_6 Z_7 Z_8 X_9 $,
        $Z_1 Z_7 $.
\end{itemize}
Any other operator from the groups in Tabs.~\ref{tab:group1_2} and \ref{tab:group3_4} can be written as a product from some operators in the lists.

In the second step, we construct the Clifford circuits, which allow the simultaneous measurement of the basis operators following Ref.~\cite{gokhale2019minimizing}.
Effectively, each basis operator from the basis is mapped to a Pauli $Z$ on a specific qubit.
Accordingly, the expectation value of products of these basis operators can then be obtained by measuring the expectation value of a Pauli string containing more than one Pauli $Z$.

\section{Zero-Noise-Extrapolation for measurement patterns}
\label{sec:error_mit}
Zero-Noise-Extrapolation (ZNE) is an error-mitigation strategy, in which a given quantum circuit is artificially stretched to amplify the noise~\cite{PhysRevLett.119.180509}.
In order to mitigate expectation values, one first measures the observable at different noise amplification levels $\lambda$ and performs an extrapolation to the zero-noise limit.

In our measurement-based approach we can use zero-noise mitigation directly on the final observable.
Following Ref.~\cite{dehaene2003clifford}, the state before measurement of the ancilla and main qubits can be written as:

\begin{equation}
    \ket{\psi} = \frac{1}{\sqrt{N}} \sum_n c_n \left(\prod_{i\in \mathcal{C}} \prod_{j\in \mathcal{N}_i} CP_{ji}\right) \ket{\psi_a}_n\ket{\psi_q}_n,
    \label{eq:State}
\end{equation}
where $c_n \in \{\pm 1, \pm i\}$ and $N$ is the number of classical measurement outcomes, $\ket{\psi_q}_n$ denotes the $n$-th possible computational state of the main register, $\ket{\psi_a}_n$ corresponds to the quantum state of the ancilla circuit, $\mathcal{C}$ is the set of main qubits, $\mathcal{N}_i$ is the set of ancilla qubits connected to the $i$-th main qubit, and $CP_{ji}$ denotes the entangling Pauli-gates between both states, that can also be performed by classical post-processing the measurement outcomes. Note that the controlled Pauli gates $CP_{ji}$ are thus always controlled by the ancilla qubits.

Without loss of generality, we may assume that only $Z$-expecation values have to be measured on the main register (any basis change can be absorbed into the definition of the state in Eq.~\eqref{eq:State}). Let $\mathcal{Z}_q\in \{I, Z\}^{N}$, then:
\begin{equation}
    \braket{\psi|\mathcal{Z}_q |\psi} = \frac{1}{N} \sum_{n,m} c_m^* c_n \prescript{}{m}{\bra{\psi_a}}\prescript{}{m}{\bra{\psi_q}} \left(\prod_{i\in \mathcal{C}}\prod_{j \in \mathcal{N}_i} CP_{ji} \; \mathcal{Z}_q \; CP_{ji}\right) \ket{\psi_a}_n\ket{\psi_q}_n.
    \label{eq:ExpectationValue}
\end{equation}
Using the properties of the Clifford group (cf.~Eq.~\eqref{eq:ShiftingRules2}), we may drag the Pauli corrections $CP_{ji}$ across the Pauli string $\mathcal{Z}_q$:
\begin{equation}
    CX_{ji}(I_j\otimes Z_i) CX_{ji} = Z_j\otimes Z_i.
\end{equation}
Consequently, we may rewrite
\begin{equation}
    \prod_{i\in \mathcal{C}} \prod_{j\in \mathcal{N}_i} CP_{ji} \mathcal{Z}_q CP_{ji} =\mathcal{Z}_a \otimes \mathcal{Z}_q,
    \label{eq:OperatorGlobalisation}
\end{equation}
where $\mathcal{Z}_a$ is the Pauli string acting on the ancilla register.
This step effectively transforms the Pauli string $\mathcal{Z}_c$, which was previously only acting on the main register, to an operator, that acts also on the ancilla register. 
By inserting Eq.~\eqref{eq:OperatorGlobalisation} into Eq.~\eqref{eq:ExpectationValue}, we obtain
\begin{equation}
    \braket{\psi|\mathcal{Z}_q|\psi} = \frac{1}{N}\sum_{n,m} c_m^* c_n
    \prescript{}{m}{\bra{\psi_a}}\mathcal{Z}_a \ket{\psi_a}_n \prescript{}{m}{\bra{\psi_q}}  \mathcal{Z}_q \ket{\psi_q}_n.\label{eq:zne_sum}
\end{equation}
Next, we can exploit that the computational states $\ket{\psi_q}_n$ are eigenstates of the Pauli-$Z$ operator, hence

\begin{equation}
    \prescript{}{m}{\bra{\psi_q}}  \mathcal{Z}_q \ket{\psi_q}_n = (-1)^{s_n} \delta_{nm},
\end{equation}
where $s_n \in \{0, 1\}$ ensures the correct phase according to the bitstring. This leads us to the final expression 
\begin{equation}
     \braket{\psi|\mathcal{Z}_q|\psi} = \frac{1}{N}\sum_{n} (-1)^{s_n}
    \prescript{}{n}{\bra{\psi_a}}\mathcal{Z}_a \ket{\psi_a}_n.
    \label{eq:zne_sum_final}
\end{equation}
Eq.~\eqref{eq:zne_sum_final} shows, that the expectation value of an observable $\mathcal{Z}_q$ on the main qubits can be written as a sum of $Z$ expectation values on the ancilla qubits. 
The relationship between $\braket{\mathcal{Z}_q}$ and the sum over $\braket{\mathcal{Z}_a}$ shows, why zero noise extrapolation works in our approach. From Eq.~\eqref{eq:zne_sum_final} we see that amplifying the noise in the $\ket{\psi_a}_n$ states will increase the noise in the expectation values given by $\prescript{}{n}{\bra{\psi_a}}\mathcal{Z}_a \ket{\psi_a}_n$.
Since all these states are equivalent up to local unitaries for all $n$, we can increase the noise of $\braket{\mathcal{Z}_q}$ in a well-defined way by amplifying the noise in the circuits which prepare the ancilla states $\ket{\psi_a}_n$.

\section{Data acquisition for the VQE experiment}
\label{sec:data_vqe}

\begin{figure}[b!]
     \centering
    \includegraphics[width=\textwidth]{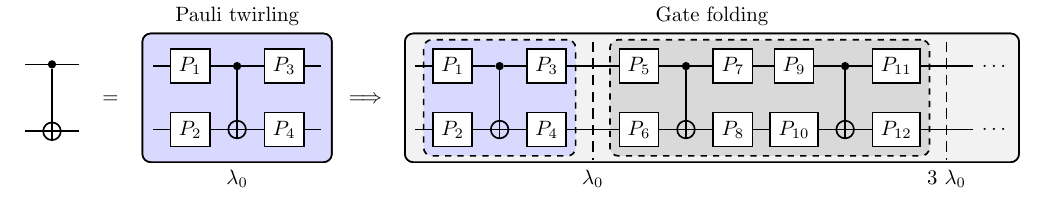}
    \hfill
    \includegraphics[width=0.4\textwidth]{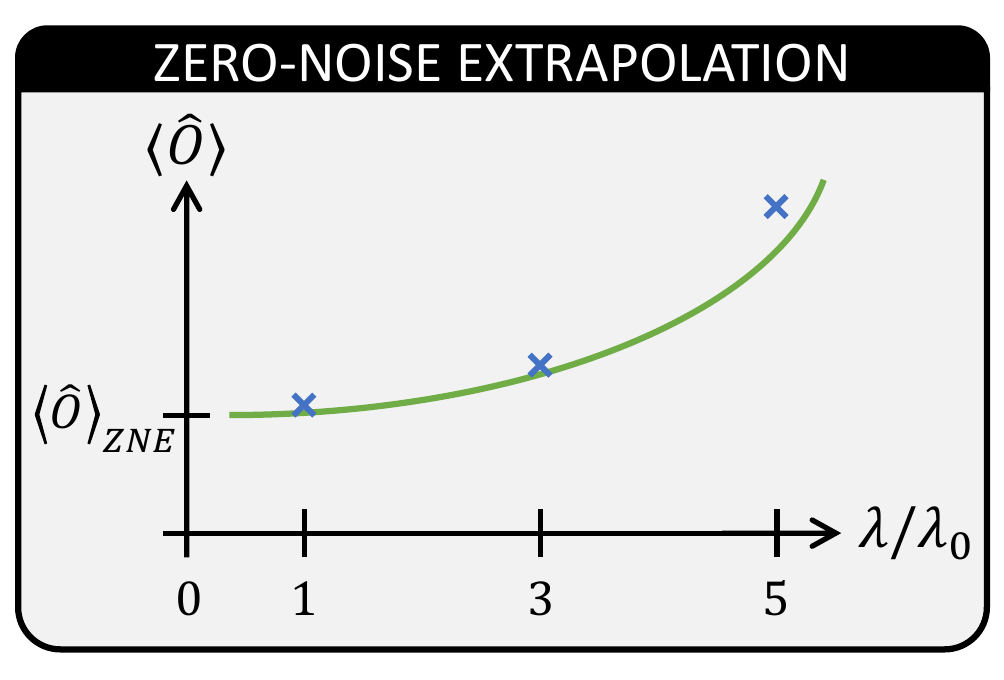}
    \caption{Overview of the error-mitigation strategies we performed. Upper panel: in order to amplify the noise, we can substitute one CNOT-gate by three. In addition, we use randomized compiling over the Pauli group (a.k.a~Pauli twirling) to mitigate coherent noise~\cite{PhysRevA.94.052325}. Lower panel: After measuring the observable at different noise levels, one can perform an extrapolation the zero-noise limit.}
    \label{fig:ZNE}
\end{figure}

In this Appendix, we report how we performed the ZNE for the case of the VQE from Sec.~\ref{sec:H2O_experiment}.
In order to increase the noise in the ancilla circuits, we use the method of local gate folding \cite{giurgica2020digital}.
In this method, all CNOT-gates in the circuit are replaced as
\begin{equation}
    CX_{ij} \rightarrow CX_{ij}^{2n+1}
\end{equation}
where $n$ is an integer.
The additional CNOT-gates do not change the outcome of the circuit, but stretch the pulse which is executed on the hardware.
This leads to a higher error-rate due to decoherence effects in the qubits, which is the main source of errors.
In order to perform ZNE, we estimate $\lambda$ by calculating the factor by which the pulse is stretched in time, see Fig.~\ref{fig:ZNE}.

In addition to ZNE we use Pauli twirling~\cite{PhysRevA.94.052325} and dynamical decoupling~\cite{PhysRevA.58.2733} as additional mitigation techniques.
For the Pauli twirling, we substitute each CNOT-gate in a given circuit randomly by
\begin{equation}
    CX_{ij} \rightarrow {P_1}_i {P_2}_j CX_{ij} {P_3}_i {P_4}_j,
\end{equation}
where $P_i=X,Z,Y$ or $I$, and where $P_1$ and $P_2$ are chosen randomly and $P_3$ and $P_4$ such, that the circuit do not change the effect of the original gate.
In total, there are $16$ different combinations how to substitute the CNOT-gate.

\begin{figure}
    \centering
    \includegraphics[width=0.99\textwidth]{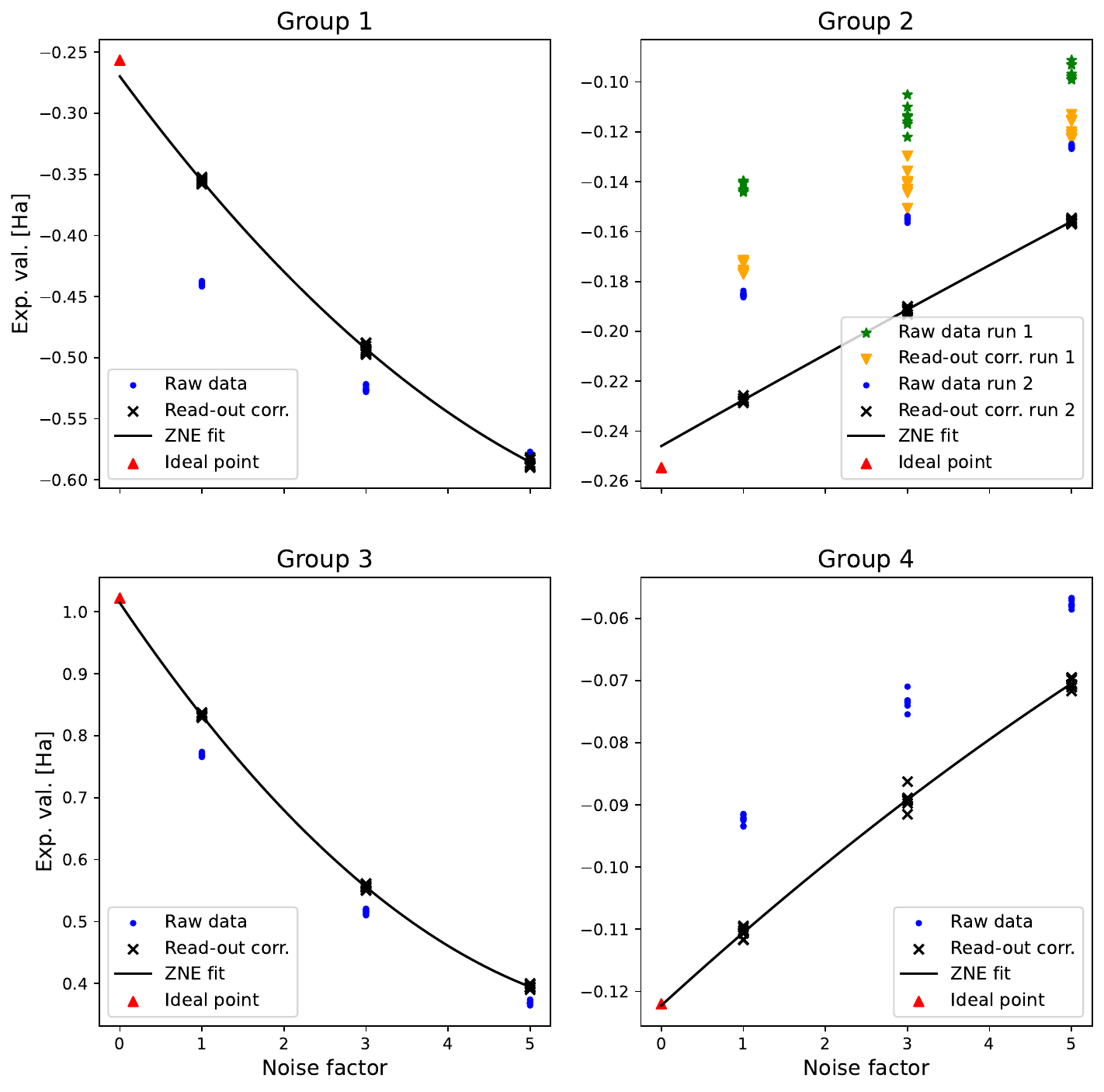}
    \caption{Data acquisition for the ZNE for measuring the expectation values of the first four groups of the partitioning of the H\textsubscript{2}O Hamiltonian. We ran each experiment eight times to see the statistical effect of shot noise. We performed read-out error mitigation using the \texttt{M3} package. In all cases, we find that read-out error mitigation together with ZNE significantly improve the extracted expectation value comparing with the ideal ones (red triangles in the plot). We performed the measurement of the expectation value of the second group twice, since the first run gave inconsistent results. This can be seen in the data points which are distributed with a much higher standard deviation compared to the data of the other runs. The second run gave much better results, although it ran just roughly $60$ minutes later.}
    \label{fig:H2O_zne}
\end{figure}

In Fig.~\ref{fig:H2O_zne} we show the results from running our circuits on \texttt{ibm\_hanoi}.
Before running the graph state circuits, we first ran read-out mitigation circuits using the \texttt{m3} package using $100\,000$ shots per circuit.
By calculating the readout-calibration matrix, we are then able to perform readout-error mitigation on the results.
As can be seen from Fig.~\ref{fig:H2O_zne}, read-out errors are an important source of error in our algorithm and need to be corrected before performing ZNE.

We evaluate the expectation values at three different noise levels, $\lambda\in \{1,3,5\}$.
In order to extrapolate to the zero-noise limit at $\lambda = 0$ we use a second-order polynomial fit:
\begin{equation}
    \braket{H(\lambda)} = a \lambda^2 + b \lambda +c,
\end{equation}
such that in the zero-noise limit the expectation value is given by $c$ and its error by the fitting error.
We ran each experiment eight times to quantify the effect of statistical shot noise.

\end{document}